\documentclass[aip,graphicx]{revtex4-1}
\usepackage{graphicx}
\usepackage{dcolumn}
\usepackage{bm}

\usepackage[utf8]{inputenc}
\usepackage[T1]{fontenc}
\usepackage{mathptmx}
\usepackage{etoolbox}
\usepackage{multirow}
\usepackage{hyperref}
\usepackage[capitalise]{cleveref}
\usepackage{caption}
\usepackage{subcaption}

\draft 

\begin{document}


\title{Study of selected mild steels for application in vacuum systems of future gravitational wave detectors} 



\author{Carlo Scarcia}
\altaffiliation{RWTH Aachen, Institute III B}
\email[email address: ]{carlo.scarcia@cern.ch}
\author{Giuseppe Bregliozzi}
\author{Paolo Chiggiato}
\author{Alice Ingrid Michet}
\author{Ana Teresa Perez Fontenla}
\author{Martino Rimoldi}
\author{Mauro Taborelli}
\author{Ivo Wevers}
\affiliation{CERN}

\date{\today}

\begin{abstract}
Next-generation gravitational wave detectors (GWDs) like the Cosmic Explorer and Einstein Telescope require extensive vacuum tubing, necessitating cost-effective materials. This study explores the viability of mild steel as an alternative to austenitic stainless steel for UHV beampipes, focusing on outgassing rates and surface chemistry after low-temperature bakeouts. Mild steels exhibit significantly lower hydrogen outgassing rates, below 10$^{-14}$ mbar l s$^{-1}$ cm$^{-2}$ after bakeouts at 80°C for 48 hours. While water vapor is the primary residual gas after such low-temperature bakeouts, repeated treatments reduce its outgassing rate and modify surface conditions so that such benefit is preserved after at least six months of exposure to laboratory air. These findings position mild steel as an economical and efficient material for future GWD beampipes. 
\end{abstract}


\maketitle 

\section{\label{sec:level1} Introduction}
Since the first recorded signals in 2015 \cite{firstdetection}, the detection of gravitational waves has opened a new window of observation into the universe, contributing to a deeper understanding of cosmological and astronomical events. Gravitational wave detectors (GWD) employ ultra-sensitive laser interferometry to reveal space-time deformations, meticulously studying and attenuating all noise sources \cite{noises}. One significant noise source is the residual gas present in the vacuum tubes of the interferometer. Pressures in the ultra-high vacuum (UHV) range are necessary to mitigate statistical fluctuations in the number of molecules within the volume traversed by the laser beam, thereby minimizing variations in the refractive index and photon phase shift. UHV conditions are designed to ensure that pressure-related noise remains at least an order of magnitude lower than the sum of all other sources of noise, including seismic, thermal, quantum, and Newtonian effects. 

While current GWDs (such as LIGO \cite{aLIGO}, Virgo \cite{aVirgo}, and KAGRA \cite{KAGRA}) continue to provide valuable insights through successive enhancements, their capabilities are limited, offering only a partial glimpse of the gravitational universe. To expand the potential discovery landscape, a third generation (3G) of GWDs is being proposed, notably the Cosmic Explorer (CE) in the USA and the Einstein Telescope (ET) in Europe. The CE proposal foresees two right-angle interferometers at distinct locations, featuring arms that are 10 times longer than those of LIGO. This configuration necessitates approximately 160 km of vacuum tubing with an internal diameter of 1.2 meters \cite{CE}. The ET project, on the other hand, will consist of six interferometers underground within a triangular tunnel measuring 10 km on each side. This setup requires approximately 120 km of vacuum piping with an internal diameter of 1 meter \cite{ET}.

The remarkable size of the vacuum system and its significant influence on the projected total cost call for revised designs, fabrication methods, and materials, as a mere scale-up of the second-generation GW detectors would be economically prohibitive. In this perspective, it is crucial to choose a material that ensures ultrahigh vacuum performance while also offering a lower cost than the austenitic stainless steels currently in existing gravitational wave observatories \cite{virgosteel}. 

A potential alternative to austenitic stainless steel for manufacturing UHV chambers could be low-carbon steels \cite{CE, NFSwork}. Commercially known as mild steel, it is widely used as a structural material due to its affordability and excellent mechanical properties. In vacuum technology, mild steel is commonly employed in systems operating within pressure ranges above 10$^{-6}$ mbar \cite{highvacuum,jousten}. However, its broad adoption in a wider spectrum of applications is limited by several factors. Firstly, it is prone to corrosion  \cite{spanishcorrosion}, and its production process is primarily tailored for structural applications, often neglecting the surface finishing. Furthermore, its ferromagnetic properties pose a considerable limitation to application in particle accelerators and surface science equipment. Additionally, mild steel exhibits a significant CO outgassing rate \cite{jousten} when exposed to high temperatures due to the high C content. 
In the past, only a few authors have described the outgassing rate of mild steels, resulting in discouraging results when compared to austenitic stainless steels. The water outgassing rates reported in the literature and textbooks often exhibit significant variability, sometimes differing by as much as three orders of magnitude \cite{oldmild, oldmild2}. Furthermore, the lack of detailed information about the grade, surface conditions, cleaning procedures, or thermal treatments in these reports complicates the ability to conduct a thorough comparison. Scarce information could also be found for the hydrogen outgassing rate after bakeout. Indeed, the most complete results were reported for a C15E steel (S15C, according to JIS standard) after a bakeout at 300°C for 3h \cite{oldmild3}, where mild steel showed values one order of magnitude higher than a 304L austenitic stainless steel baked at 150°C for 24h \cite{oldmild4}. 
Despite these discouraging results, a recent study has shown that mild steels could still be developed and applied in UHV environments \cite{Park1,Park2}. Indeed, using commercially available alloys, Park et al. measured a H$_{2}$ outgassing rate from mild steel that was 10 to 50 times lower than austenitic stainless steel subjected to a similar bakeout procedure (i.e., 150°C for 48h). The same work gave a qualitative picture of the vacuum performance of low-carbon steels; however, the information is limited only to H$_{2}$ and H$_{2}$O outgassing rates. Other interesting results were recently presented \cite{fedchack}. \\
This study aims to evaluate mild steel as a potential material for vacuum tubes 3G GWD. We measured the outgassing rates of well-defined mild steels for the typical gas species found in UHV environments, including H$_{2}$, H$_{2}$O, CH$_{4}$, CO, and CO$_{2}$. To ensure cost-effectiveness, we utilized only readily available off-the-shelf products and assessed the impact of low-temperature bakeouts on outgassing rates. Temperature Programmed Desorption (TPD) measurements were employed to estimate the H$_{2}$ content in the analysed samples. Furthermore, we conducted additional X-ray Photoelectron Spectroscopy (XPS) measurements to monitor the surface condition of the samples after heating at various temperatures.

\section{\label{sec:level2}EXPERIMENTAL APPARATUS AND METHODS}
\subsection{THROUGHPUT METHOD}
\begin{figure}[!h]
\includegraphics[scale=0.3]{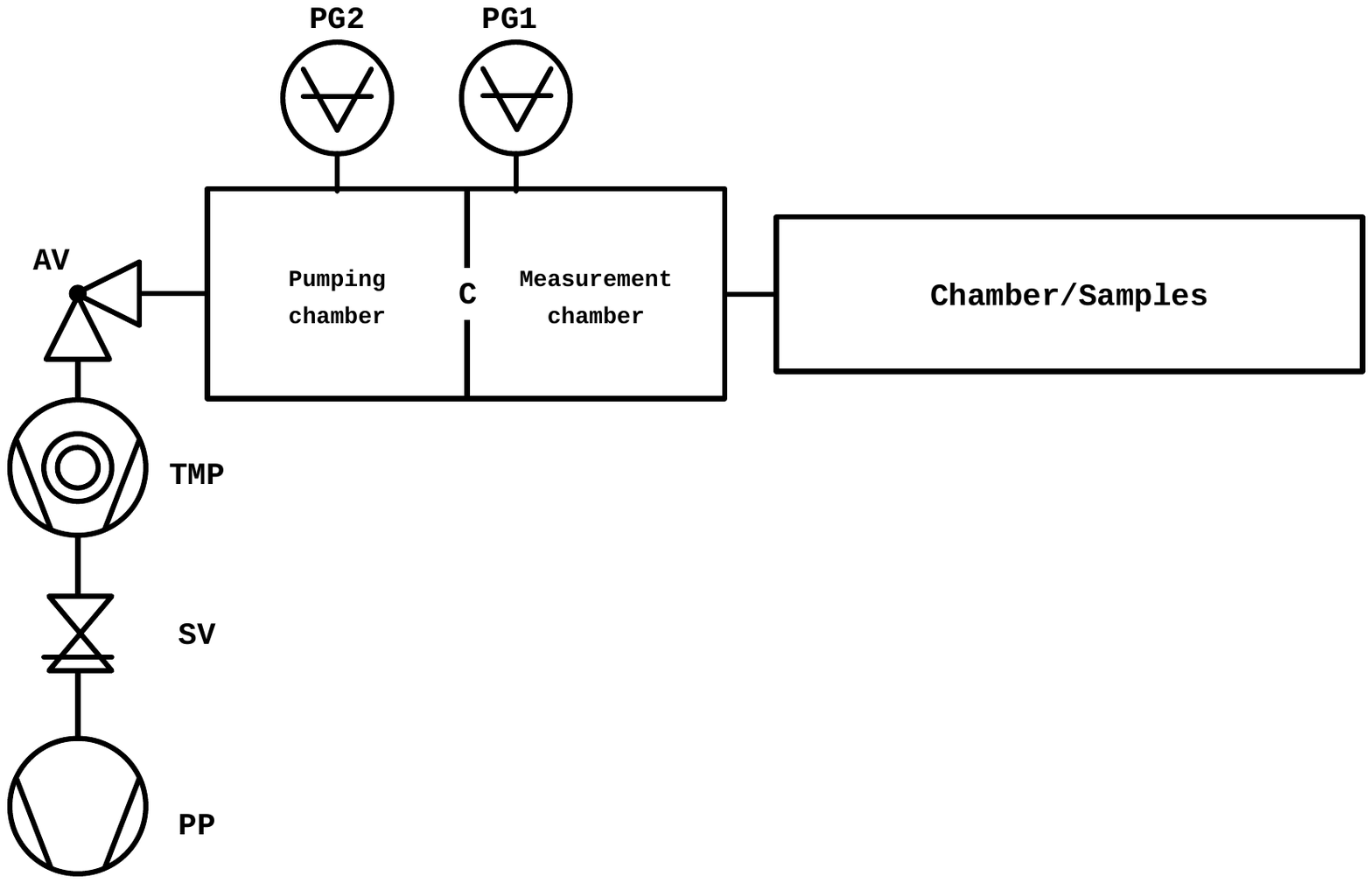}%
\caption{\label{fig:throughput}Schematic of the throughput system. TMP: turbomolecular pump; SV: electromagnetic safety valve; PP: primary pump; AV: right angle valve; C: orifice conductance; PG1 and PG2: pressure gauge 1 and 2.}%
\end{figure}

The outgassing rates of unbaked samples were measured using the throughput method \cite{redhead,chiggiato}. The measurement consisted of monitoring as a function of pumping time, the pressure in the test dome divided into two sides by the interposition of an orifice. On one side, the sample was placed either as a set of specimens inserted in a sample holder or as a connected vacuum chamber. On the other side of the orifice, a turbomolecular pumping group was connected by means of an all-metal right-angle valve (see Fig.\ref{fig:throughput}).
The outgassing rate of the sample per unit of geometrical surface area tested is calculated as:
\begin{equation}
  q_{i}  = \frac{[C_{i} \cdot (P_{1}-P_{1,BGD})-(P_{2}-P_{2,BGD})] }{A_{sample}} \;\; \bigg[\frac{mbar \cdot l}{s \cdot cm ^ {2} }\bigg]
\label{eq:one}
\end{equation}
where C$_{i}$ is the conductance of the oriﬁce for the gas of interest (index ‘i’); P indicates the measured pressures in the sample side, with index ‘1’, and pump side, with index ‘2’. The index ‘BGD’ stands for background, and it indicates the pressures recorded at the same pumping time without the sample installed in the system. The subtraction of background pressures is required to remove the contribution of the austenitic stainless steel test dome and gauges to the total outgassing rate. The geometrical surface area of the sample is indicated as A$_{sample}$. During the measurements, the temperature was stabilised at 21±2°C.
The orifice had a diameter of 0.8 cm, which results in $C_{H_2O}$ = 7.4 ls$^{-1}$. The test dome had a diameter of 10 cm; it was connected to the sample holder or the sample chamber by a DN100 CF flange. The pressure measurement was performed by cold cathode gauges (Pfeiffer IKR070, estimated accuracy ±30\%). The speciﬁc outgassing rate was monitored for approximately 100 h of pumping. The background measurements were repeated every time a new steel grade was tested.

\subsection{\label{sec:accu}COUPLED ACCUMULATION-THROUGHPUT METHOD}
\begin{figure}[!h]
\includegraphics[scale=0.5]{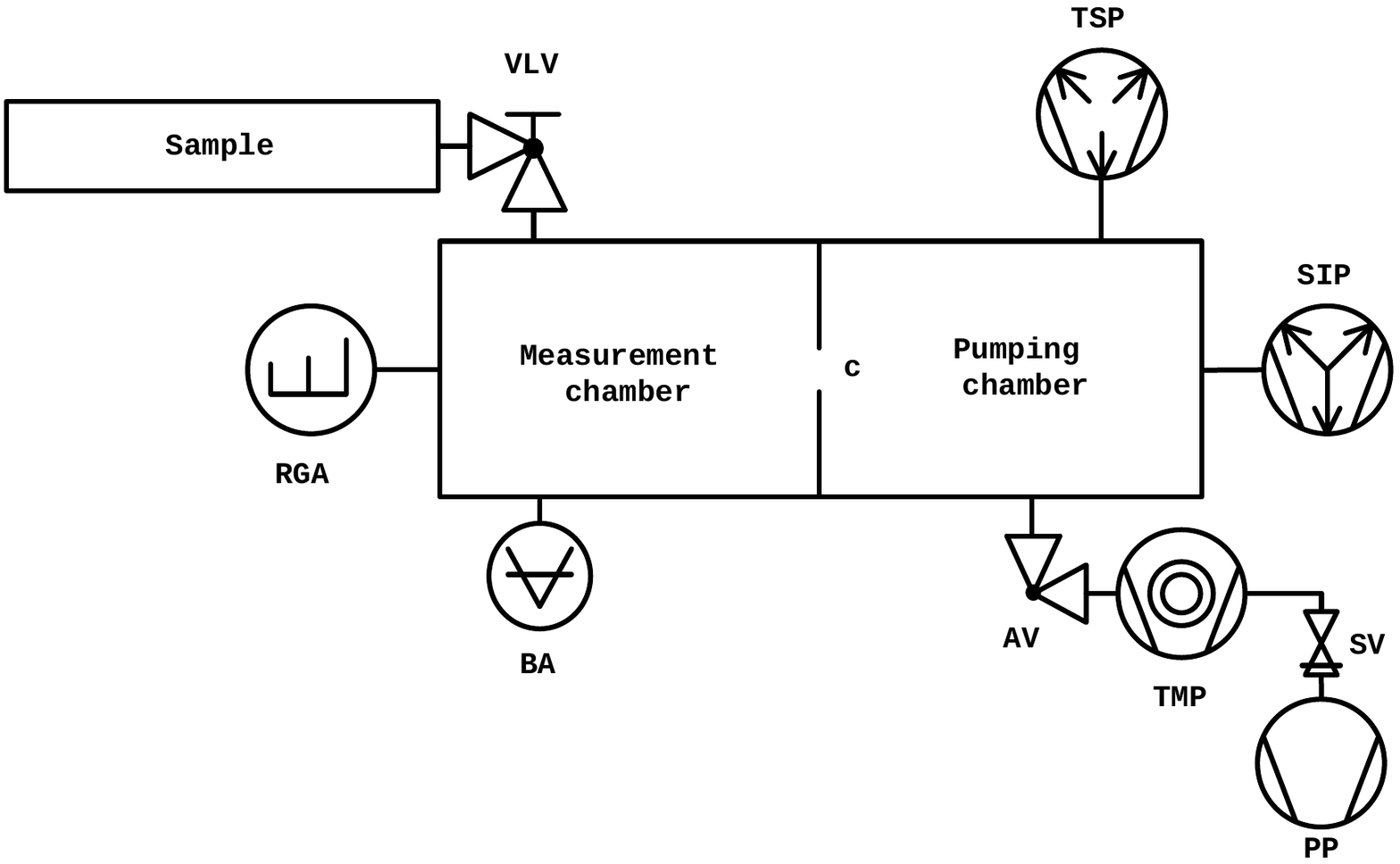}%
\caption{\label{fig:accu}Schematic of the coupled accumulation – throughput system. TMP: turbomolecular pump; SV: electromagnetic safety valve; PP: primary pump; AV: right angle valve; C: orifice conductance; BA: Bayard-Alpert gauge; SIP: sputter ion pump; TPS: Ti sublimation pump; RGA: residual gas analyser.}%
\end{figure}

The coupled accumulation-throughput method \cite{chiggiato} was employed to measure the post-bakeout outgassing rate of the samples (see Fig.\ref{fig:accu}). The all-metal measurement system consisted of a throughput system connected to the sample holder via a variable leak valve (VLV). Alternatively, the sample was the vacuum vessel to be measured connected directly to the VLV. Pumpdown, leak detection, and bakeout procedures were routinely conducted with the VLV left fully open. However, when performing an outgassing measurement, the VLV was fully closed, allowing the released gas to accumulate within the inner volume of the sample holder for a specified accumulation period. Subsequently, the VLV was gradually opened. The gas escaping from the sample holder was then detected using a residual gas analyser (RGA, Pfeiffer QMA 125) employing multiple ion detection (MID) and a Bayard-Alper gauge (BA, SVT305 CERN\cite{SVT}, accuracy ±10\%). The detection terminated when the pressure reached a reasonably stable value, and the VLV was closed for a new accumulation.  
A TMP group ensured the required pumping speed during pumpdown and bakeout. To attain H$_{2}$ pressures in the low 10$^{-11}$ mbar range, a Sputter Ion Pump (SIP) and a Titanium Sublimation Pump (TSP) were employed during the system operation. A 0.8 cm diameter orifice separated the pumping chamber from the measurement chamber where the RGA and BA gauge were installed. To avoid drift in the sensitivities, the RGA was regularly calibrated in situ against the BA gauge \cite{insitu}. 
Unlike other accumulation techniques, this method provides outgassing rate values for specific gas species, assuming their partial pressure during accumulation varies linearly with time. This condition is met for gases with negligible sticking probabilities on sample holder surfaces, typically observed with hydrogen and methane. However, water vapor, prone to re-adsorption on sample surfaces, violates this linearity. H$_{2}$O molecules readsorb on the surfaces of the sample holder to attain equilibrium between accumulated gas and surface coverage. An advantage of this system is the absence of indirect gas pumping or cracking since no ion gauges are installed in the accumulation volume. Measurements are repeated for different accumulation times to confirm the measurement's linearity and increase its accuracy.
The accumulated quantity of gas Q$_{acc}$, in the time interval t$_a$, is calculated using Eq.\ref{eq:accu}.
\begin{equation}
    Q_{acc} = S_{c} \cdot \int_{t_{a}}^{t_{a} + \Delta t} I_{RGA}(\tau) \times \alpha_{RGA} \,d(\tau) \;\;\bigg[\frac{mbar \cdot l}{s}\bigg] \label{eq:accu}   
\end{equation}
where $\Delta t$ is the actual duration of the RGA recording, $I_{RGA}$ is the ion current read by the RGA, $\alpha_{RGA}$ is the calibration factor relating current and pressure for the gas of interest, and $S_{C}$ denotes the effective pumping speed at the level of the RGA. Assuming linearity, the specific outgassing rate q of the sample is computed as:
\begin{equation}
    q_{acc} = \frac{Q_{acc} - Q_{BGD}}{A_{s}} \;\;\bigg[\frac{mbar \cdot l}{s \cdot cm ^ {2} }\bigg] \label{eq:accu1}   
\end{equation}
$Q_{BGD}$ is the accumulated gas quantity measured without samples (background) after undergoing an identical bakeout cycle. If the sample is a vacuum vessel, $Q_{BGD}$ accounts for the gas released from the VLV and flanges used for vessel closure.
The sample's geometrical surface area was maximized to increase system sensitivity. To minimize other gas sources, all components that constitute the throughput system and sample holder were vacuum fired for 2 h at 950°C \cite{chiggiato} prior to installation. The VLV was dismounted, and each stainless steel sub-component underwent the same vacuum firing treatment.
\\
The sample holder was baked at 80°C and 150°C for 48 h. The throughput system and VLV underwent bakeout at temperatures ranging from 200°C to 350°C. The accumulation measurements started when the samples were at room temperature (21±2°C), 24 h after the end of the bakeout cycle. 
\subsection{TEMPERATURE PROGRAMMED DESORPTION}
Temperature Programmed Desorption (TPD) measurements were conducted to ascertain the diffusible hydrogen content in the mild steel samples. The TPD analysis was carried out using a commercial TPD workstation\cite{TPDhiden}.
The system consisted of a testing chamber and a load-lock chamber, which were separated by a manual gate valve (GV). This GV maintained a UHV environment in the testing chamber, with a base pressure of approximately 1.5$\times$10$^{-9}$ mbar, while preserving the high vacuum conditions in the load-lock chamber, with a base pressure of about 1.5$\times$10$^{-7}$ mbar. This setup obviated the need for a bakeout of the test chamber upon inserting a new sample.  Cold cathode gauges were installed in both chambers to monitor pressure levels. Additionally, the testing chamber featured an RGA (Hiden 3F PIC) directly facing the sample to monitor gas evolution during measurements.
The samples, each having a surface area of 2 cm² and thickness varying from 0.069 to 0.4 cm, underwent heating from 25°C to 940°C at a ramp rate of 5°C/min. To ensure measurement reproducibility, the background was re-measured after every ten samples. The accuracy of quantitative measurements was verified through regular in-situ calibration of the RGA. Hydrogen concentration calculation relied on sample weight, measured with a weight scale with a sensitivity of ±0.1 mg.
\subsection{X-RAY PHOTOELECTRON SPECTROSCOPY}
The surface chemical composition and chemical states were characterized by X-ray Photoelectron Spectroscopy (XPS) using a commercially available UHV system (measurement chamber base pressure \textless 3$\times$10$^{-10}$ mbar) from SPECS Surface Nano Analysis GmbH equipped with a monochromatized Al K$\alpha$ source. The energy scale of the electron analyser was calibrated on the Au 4f$_{7/2}$ and Cu 2p$_{3/2}$ lines of sputter-cleaned samples. 
The 1 cm2 samples were measured at normal emission at room temperature. XPS spectra were taken after in-situ thermal treatment at 80°C and 150°C (heating ramp of 3°C/min, steady state of 17 h) to investigate the surface evolution during bakeout.
\subsection{E.	ULTIMATE PRESSURE AFTER LOW-TEMPERATURE BAKEOUT}
Ultimate pressures after low-temperature bakeouts were investigated using a dedicated system. The sample, a mild steel vacuum chamber, underwent a series of consecutive bakeouts at 80°C, each lasting 48 h, until the system's pressure limit was attained. The ultimate pressure at room temperature (21±2°C) was measured 24 h after the end of each bakeout, without intermediate air venting.
\begin{figure}[!h]
\includegraphics[scale=0.5]{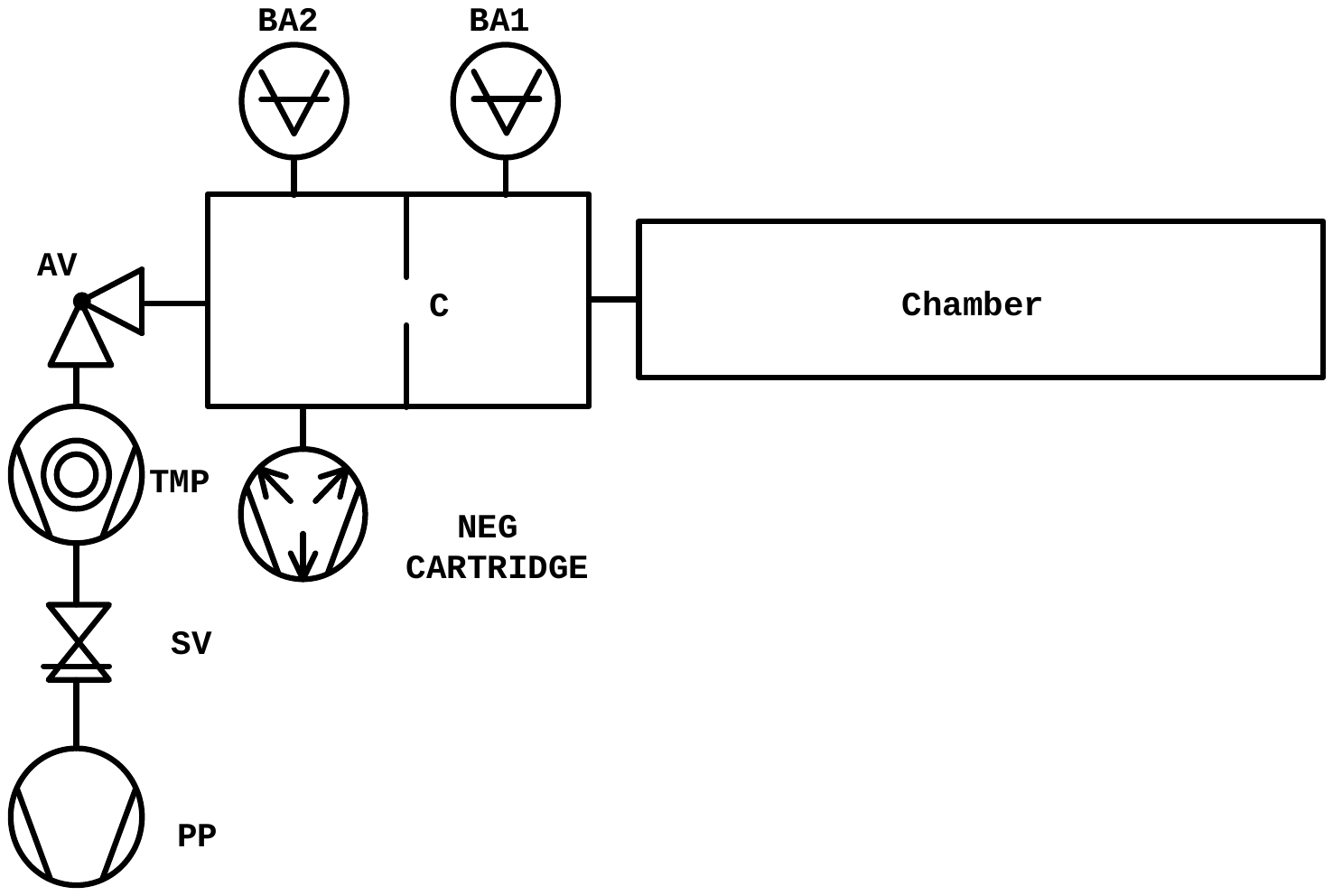}%
\caption{\label{fig:UPS}Schematic of the ultimate pressure system. NEG: Non Evaporable Getter.}%
\end{figure}\\
All the stainless-steel components of the system (see Fig.\ref{fig:UPS}) were vacuum fired at 950°C for 2h to reduce their hydrogen outgassing rate. The sample was pumped through an orifice (C) with a diameter of 1 cm (9.2 ls$^{-1}$ for N$_{2}$) by a Non-Evaporable Getter (NEG) cartridge that provided a 2000 ls$^{-1}$ nominal pumping speed for H$_{2}$. A TMP group (effective pumping speed 120 ls$^{-1}$ for N$_{2}$) was also installed to ensure pumping during bakeouts and to remove species that are not adsorbed by the NEG pump, i.e. rare gases and methane. The pressure measurements across the orifice were carried out by calibrated BA gauges (SVT305 CERN, accuracy ±10\%). The mild steel sample was connected to the test system by a DN100 CF AISI 316LN vacuum fired flange welded at the extremity. During the bakeout of the sample, the rest of the system was baked at temperatures in the range 200°C to 350°C, keeping the vacuum gauges always at the highest temperature. If the sample is not installed, the achieved pressure after bakeout in the dome where BA2 is installed is about 2$\times$$10^{-12}$ mbar (N$_{2}$ equivalent).
\section{MATERIAL SELECTION AND SAMPLE PREPARATION}
Mild steel off-the-shelf blocks, sheets/plates, and tubes were provided by different suppliers. They were compared with AISI 304L stainless steel available at CERN and meant for UHV application. The characteristics of the selected steels are listed in Tab. \ref{tab:material}. 
Samples for the analysis were cut from blocks and plates to the required dimensions using a bench shear. Before sample cutting, the blocks were milled on the external surfaces to remove the surface damaged layer.
In the case of tubes, DN100 CF AISI 316LN  flanges were welded at both ends after cutting samples for TPD and  XPS analysis. DN100 CF AISI 316LN  blank flanges were also prepared to perform pump-down, accumulation and ultimate pressure measurements. All flanges prior to welding were vacuum fired at 950°C for 2 h. The part of the blank flanges exposed to the vacuum was thinned to increase the hydrogen degassing efficiency of vacuum firing\cite{Calder}. Mild steel samples were then cleaned in a solvent bath (Dowclean™ 1601 for tubes and Topklean™ MC 20A for the other samples) by immersion, ultrasonic agitation, and, finally, dried in an air furnace at 60°C for 30 min before being tested. The AISI 304L samples, instead, were cleaned following the CERN UHV standard procedure\cite{304Lclean} that implies the use of a detergent bath (NGL 17.40 spec. ALU III), rinsing with de-mineralized water, and drying in an air furnace at 60°C for 10-60 min.
\begin{table}[h]
\caption{Chemical composition (wt.\%) of the selected steels with the corresponding manufacturing process (MP), heat treatment (HT), and shape. AISI 304L chemical composition values are to be intended as the maximum content allowed\cite{304LEDMS}. HR: Hot rolled, CR: Cold rolled, AP: Acid pickled, ERW: Electric Resistance Welded.}
\begin{tabular}{ccccccccc}
\hline
 &S355J+AR&S355J2+N&FB580& ULC-IF & \begin{tabular}[c]{@{}l@{}}ARMCO\\(grade 4)\end{tabular} &S355J2H &P355N &AISI 304L\\
\hline
MP & HR & HR + Forged & HR &CR &CR &HR+ERW &HR+AP+ERW &CR \\
\hline
HT & None & Normalization & None &None &None &None &Normalization &Solution annealed \\
\hline
Shape & Block & Block & Sheet & Sheet & Sheet	& Tube	& Tube & Plates\\
\hline
C	&0.14	&0.14	&0.086	&0.013	&0.002	&0.20	&0.19	&0.03 \\
Mn	&1.46	&1.46	&1.35	&0.099	&0.04	&1.32	&1.31	&2.0\\
Si	&0.20	&0.20	&0.045	&0.005	&0.003	&0.17	&0.16	&1.0\\
Cu	&0.06	&0.06	&0.009	&0.008	&0.008	&0.08	&0.13	&-\\
Al	&0.03	&0.03	&0.038	&0.049	&0.002	&0.03	&0.03	&-\\
S	&0.01	&0.01	&0.0012	&0.0094	&0.0018	&0.01	&0.0001	&-\\
P	&0.01	&0.01	&0.0108	&0.0093	&0.004	&0.01	&0.01	&0.03\\
N	&0.01	&0.01	&-	&-	&-	&0.004	&0.01	&0.02\\
Cr	&-	&-	&0.025	&0.023	&0.014	&-	&-	&17-20\\
Ni	&-	&-	&0.01	&0.011	&0.015	&-	&-	&10-12.5\\
Fe	&Remainder	&Remainder	&Reminder	&Reminder	&Remainder	&Remainder	&Remainder	&Remainder\\
\hline
\label{tab:material}
\end{tabular}
\end{table}
\clearpage
\section{Results and discussion}
\subsection{Water vapor outgassing rate}
The measured pumpdown curves, representing the water vapor specific outgassing rate as a function of pumping time, are shown in Figure 4. The specific outgassing rates recorded after 10 h of pumping are summarized in Table \ref{tab:pumpdown}.
\begin{figure}[!h]
\includegraphics[width=1\textwidth]{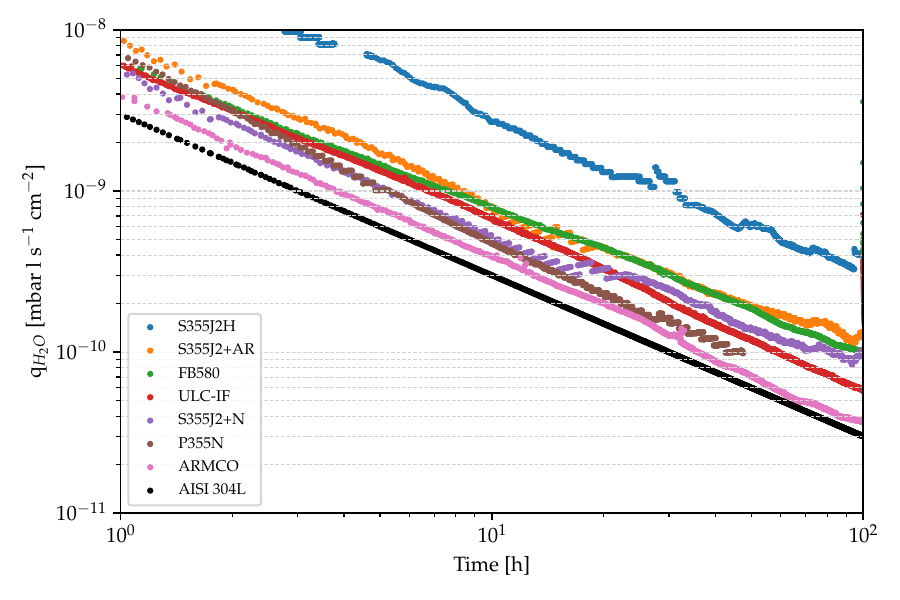}%
\caption{\label{fig:pumpdown}Pumpdown curves at 21±2°C; the specific outgassing rate is plotted as a function of the pumping time. The background value is subtracted. The 304L curve is calculated from an empirical relationship generally applied at CERN for austenitic stainless steels applied in UHV\cite{chiggiato}.}%
\end{figure}\\
As shown in Figure \ref{fig:pumpdown}, the pumpdown values of all mild steel samples are reasonably fitted by inverse power laws in which the exponents of time are around -1. This is the typical behaviour of metal surfaces discussed and interpreted by some authors\cite{fred,edwards}. In terms of quantitative data, there is a significant dispersion ranging from 1.2 to 10 times higher values than AISI 304L. The lowest specific outgassing rates are reported for samples cut from blocks, while the highest ones are recorded for the S355J2H pipe.
Scanning Electron Microscope (SEM) micrographs were taken for the S355J2H surface (see Fig.\ref{fig:tube1}). The S355J2H surface morphology appeared rough, characterized by cracks and small pores, and covered by particulates of a few $\mu$m size. Further, a SEM-FIB (Focus Ion Beam) cross-section (Fig.\ref{fig:tube2}) displays a 10 $\mu$m thick oxide layer covering the S355J2H surface. The oxide exhibits significant cracks, partial detachment from the substrate, and pores that appear open to the surface, features not seen for the other mild steel samples (see \cref{fig:tube3,fig:tube4,fig:ar1,fig:ar2,fig:n1,fig:n2} for comparison). The poor surface condition may be responsible for the high rate of water outgassing.\\
\begin{table}[h]
\caption{Water vapor specific outgassing rates measured at 21±2°C. **value estimated from the inverse power law fit at 10 h. Background removed.}
\begin{tabular}{cccc}
\hline
 Steel& \multicolumn{1}{c}{\begin{tabular}[c]{@{}c@{}}Tested samples\\ {[}cm$^{2}${]}\end{tabular}} & \multicolumn{1}{c}{\begin{tabular}[c]{@{}c@{}}q$_{10 h}$\\ {[}mbar l s$^{-1}$ cm$^{-2}${]}\end{tabular}} & q$_{0}$ $t_h^{-a}$\\
\hline
S355J2H	& 8821 & 2.7$\times$10$^{-9}$ & 2.6$\times$10$^{-8}$\;$t_h^{-0.96}$\\
S355J2+AR & 3536 & 8.1$\times$10$^{-10}$ & 8.0$\times$10$^{-9}$\;$t_h^{-0.99}$\\
FB580 & 5956 & 7.9$\times$10$^{-10}$ & 6.9$\times$10$^{-9}$\;$t_h^{-0.95}$\\
ULC-IF & 6092 & 6.7$\times$10$^{-10}$ & 6.2$\times$10$^{-9}$\;$t_h^{-0.97}$\\
S355J2+N & 5780 & 4.9$\times$10$^{-10}$ & 5.1$\times$10$^{-9}$\;$t_h^{-1.02}$\\
P355N & 5938 & 4.7$\times$10$^{-10}$ & 6.8$\times$10$^{-9}$\;$t_h^{-1.16}$\\
ARMCO & 6050 & 3.9$\times$10$^{-10}$ & 4.0$\times$10$^{-9}$\;$t_h^{-1.01}$\\
AISI 304L & - & 3.0$\times$10$^{-10}$$^{**}$ & --\\
\hline
\label{tab:pumpdown}
\end{tabular}
\end{table}
\begin{figure}[!h]
\includegraphics[scale=0.4]{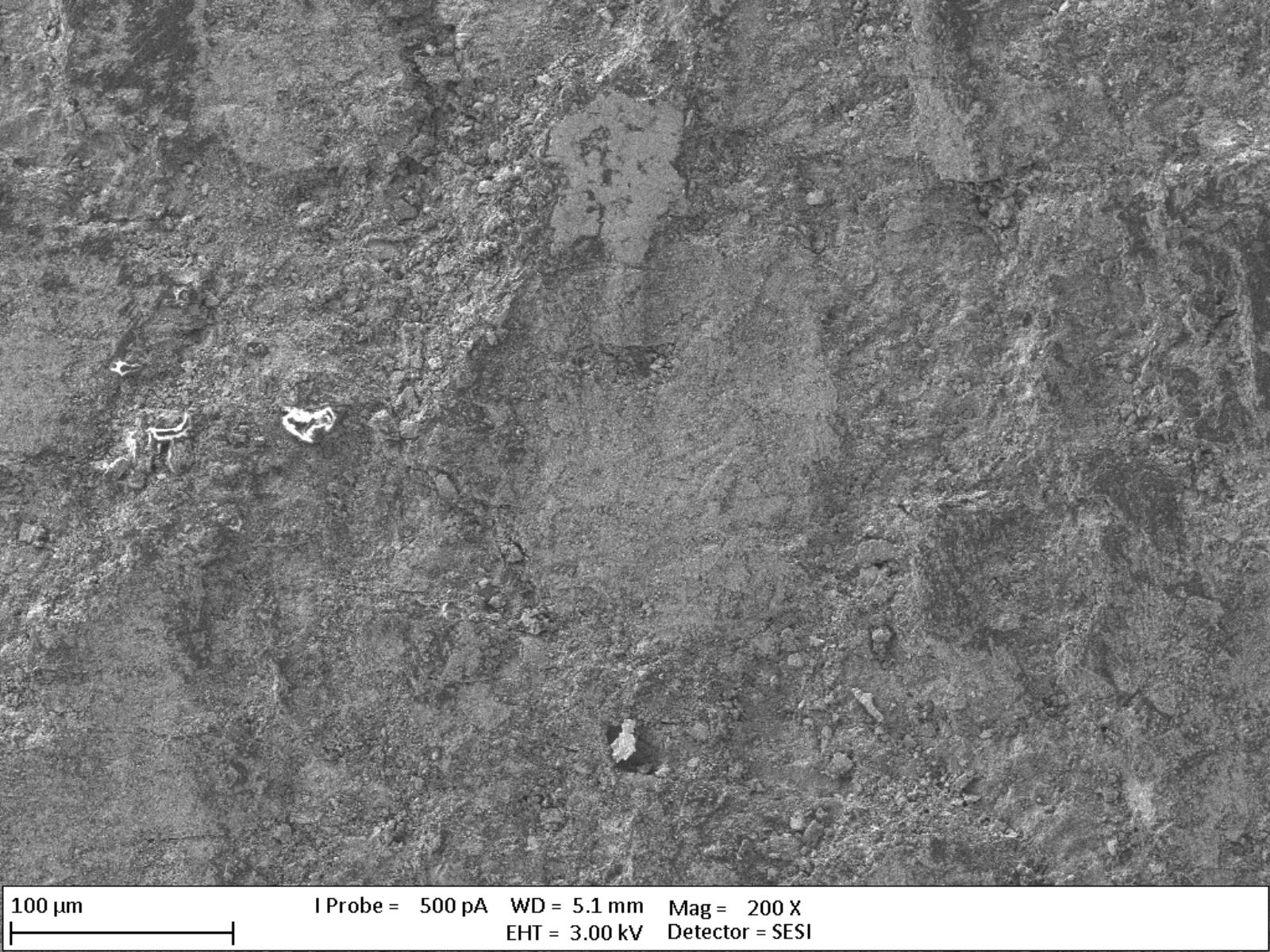}%
\caption{\label{fig:tube44}SEM micrograph of the surface of the S355J2H sample.}%
\end{figure}

\begin{figure}[!h]
\includegraphics[scale=0.4]{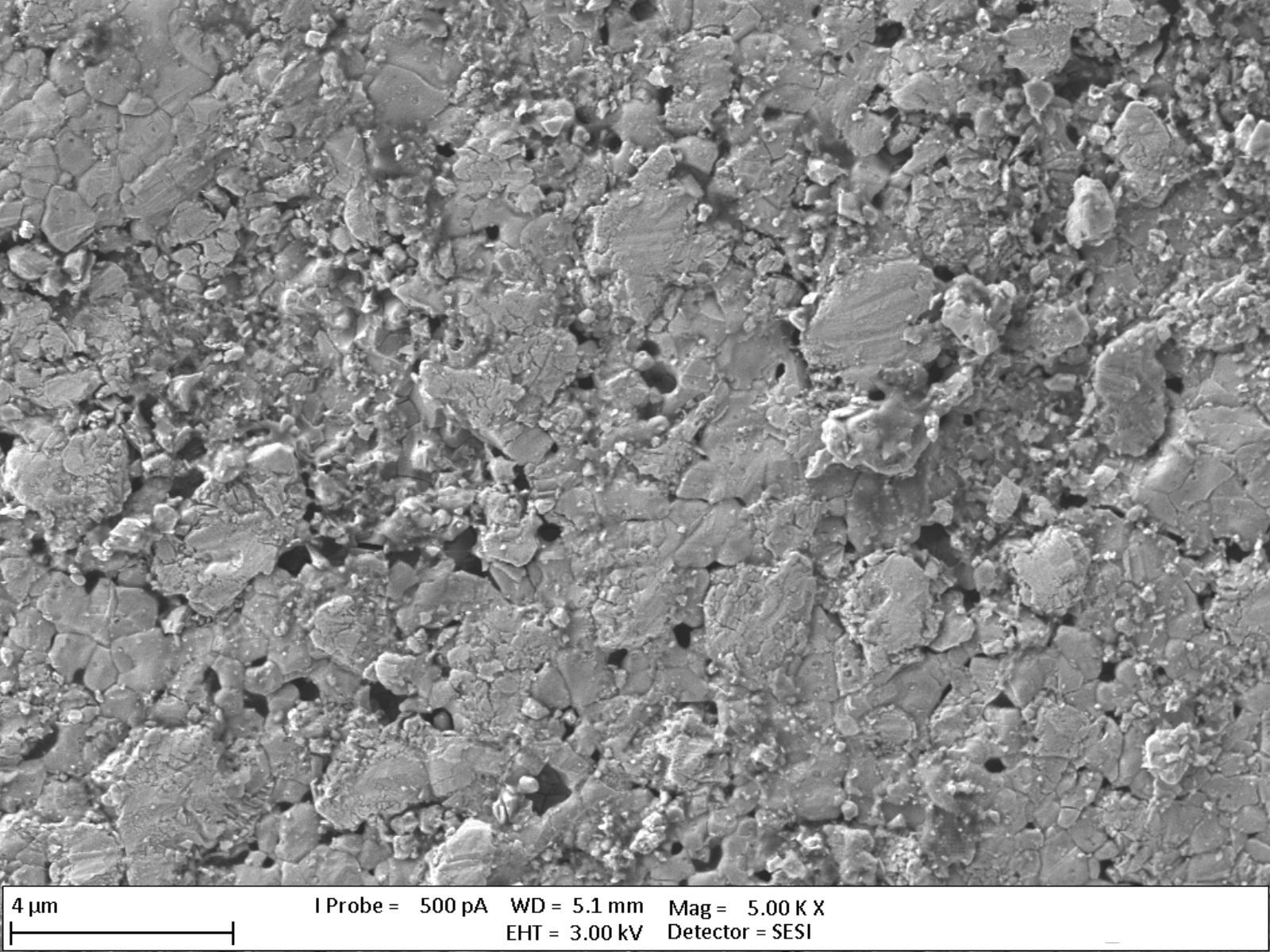}%
\caption{\label{fig:tube1}SEM micrograph of the surface of the S355J2H sample. Pores and flakes are covering the surface.}%
\end{figure}
\begin{figure}[!h]
\includegraphics[scale=0.4]{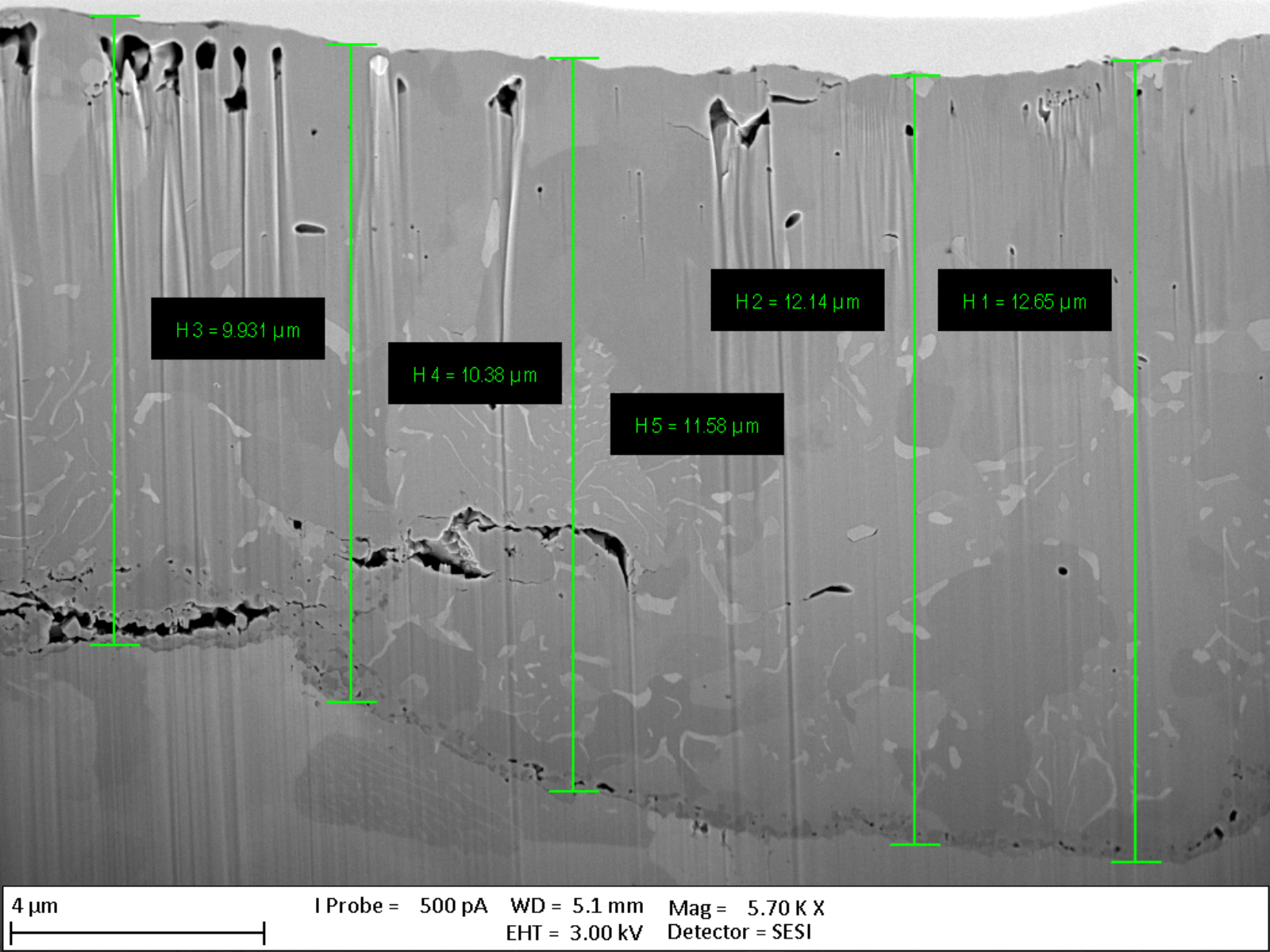}%
\caption{\label{fig:tube2}SEM-FIB cross-section micrograph of the S355J2H surface. The thickness of the oxide layer is highlighted by bars in five different positions.}%
\end{figure}
\begin{figure}[!h]
\includegraphics[scale=0.4]{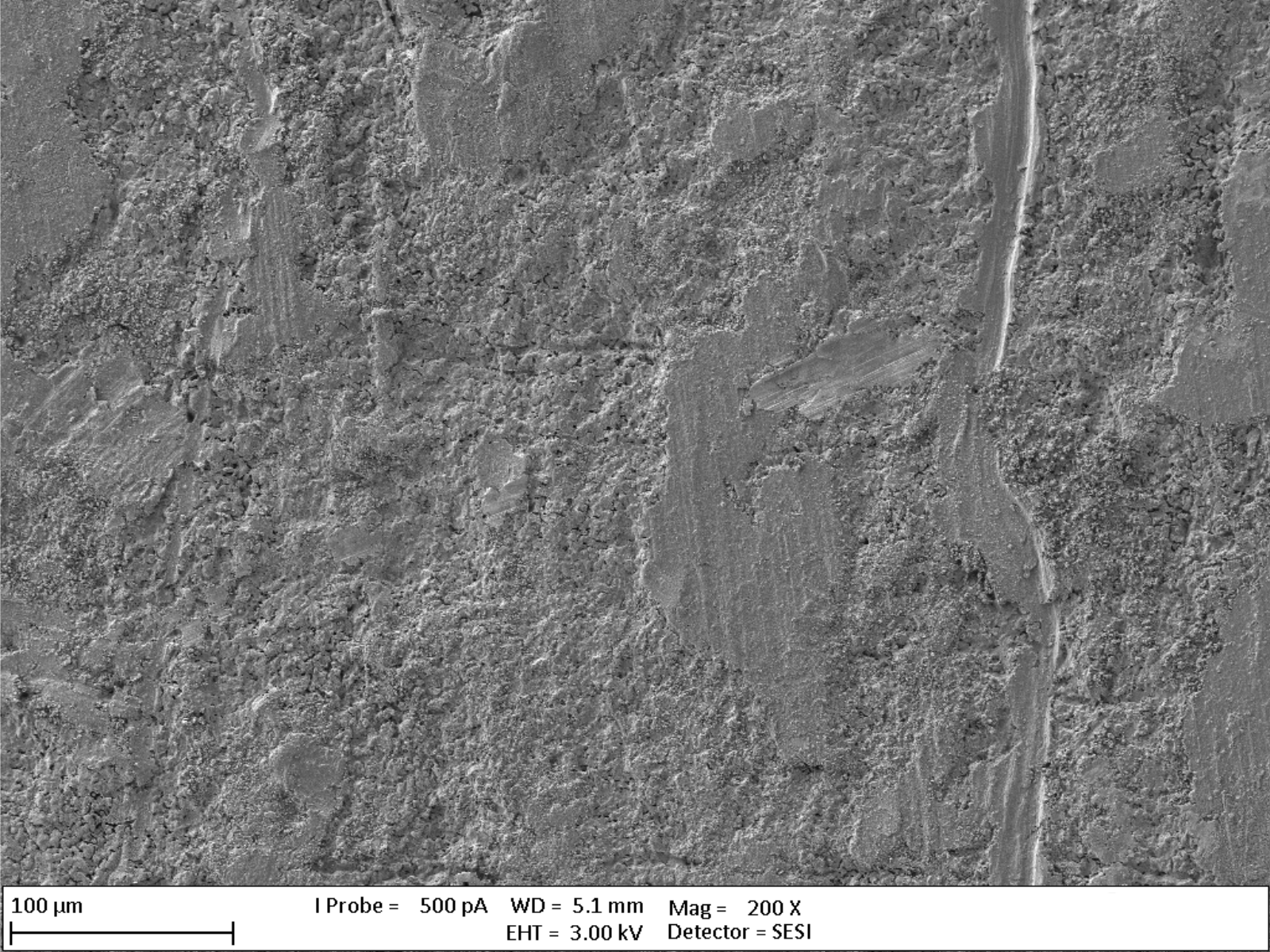}%
\caption{\label{fig:tube3}SEM micrograph of the surface of the P355N sample.}%
\end{figure}
\begin{figure}[!h]
\includegraphics[scale=0.4]{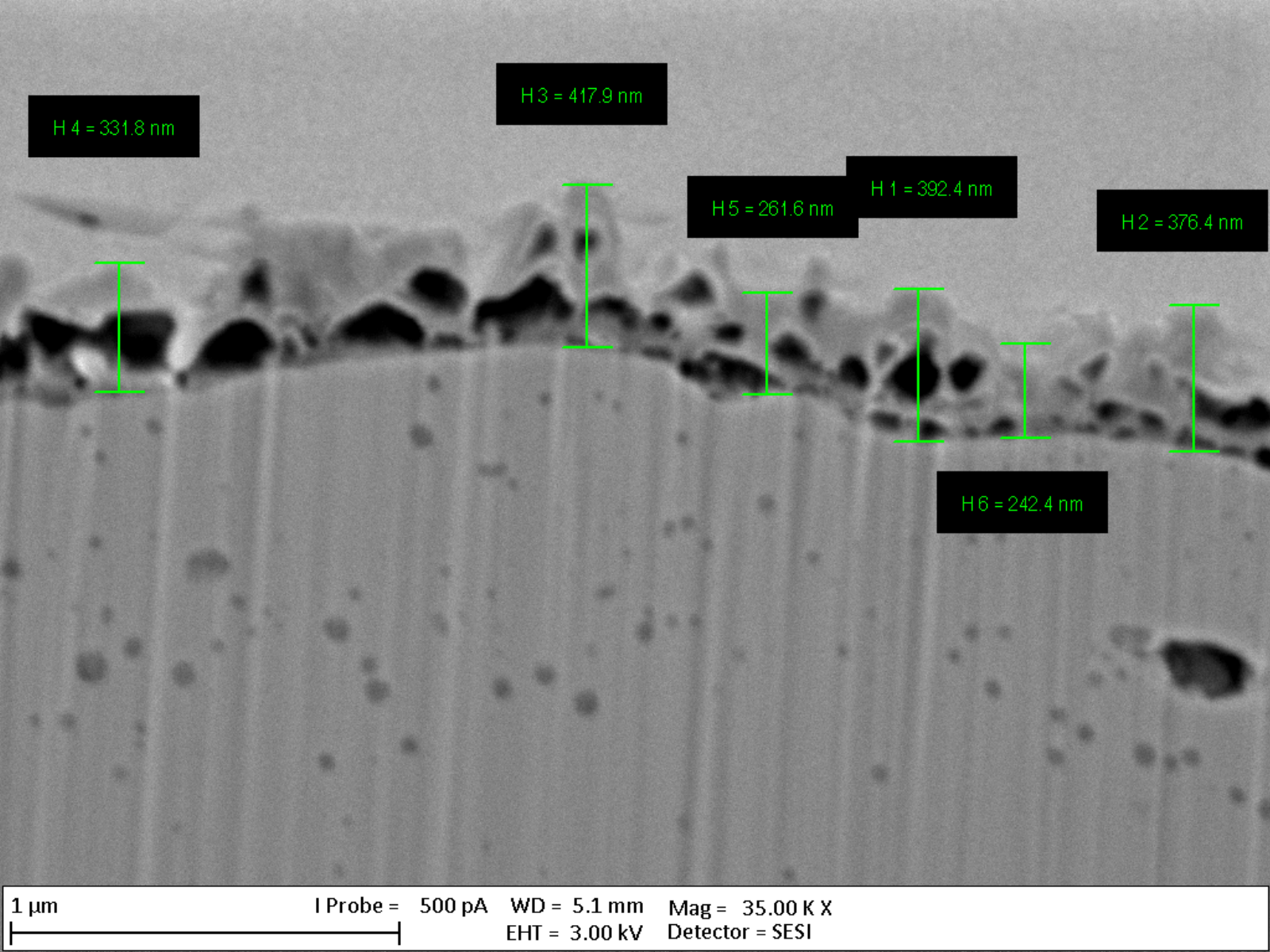}%
\caption{\label{fig:tube4}SEM-FIB cross-section micrograph of the P355N surface highlighting an oxidated spot.The thickness of the oxide layer is highlighted by bars in six different positions.}%
\end{figure}
\begin{figure}[!h]
\includegraphics[scale=0.4]{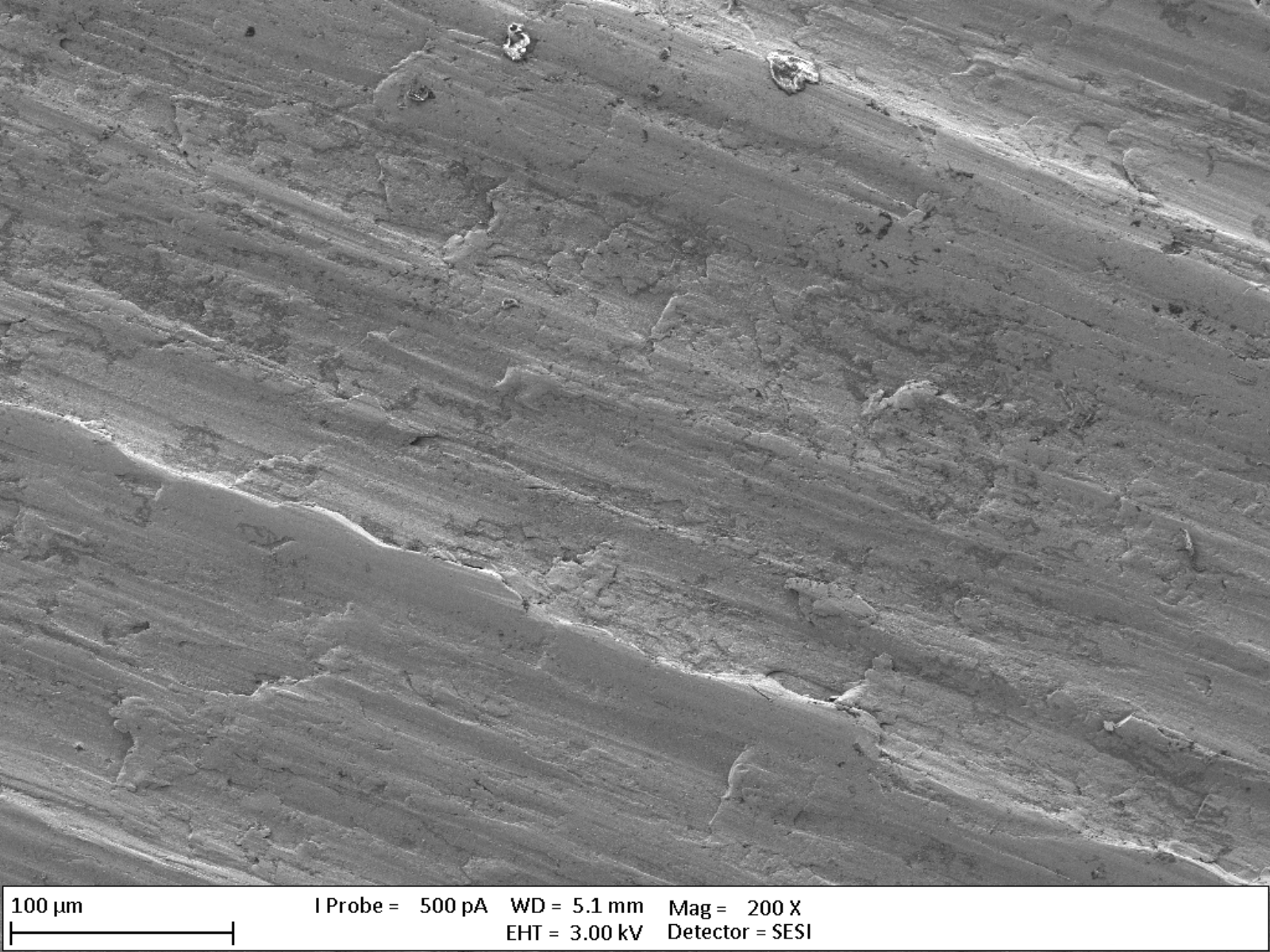}%
\caption{\label{fig:ar1}SEM micrograph of the surface of the S355J2+AR sample.}%
\end{figure}
\begin{figure}[!h]
\includegraphics[scale=0.4]{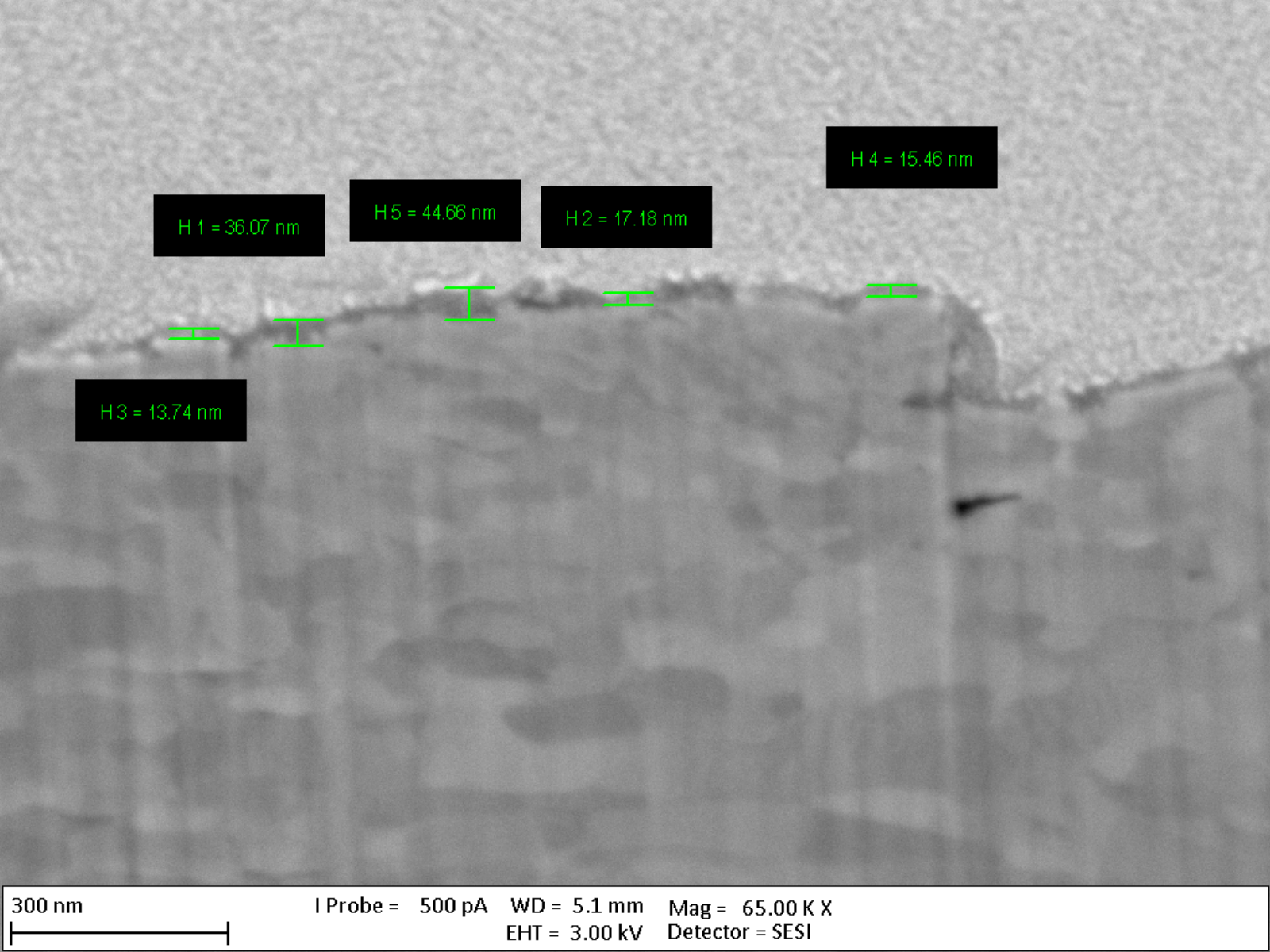}%
\caption{\label{fig:ar2}SEM-FIB cross-section micrograph of the S355J2+AR surface highlighting an oxidated spot.The thickness of the oxide layer is highlighted by bars in five different positions.}%
\end{figure}
\begin{figure}[!h]
\includegraphics[scale=0.4]{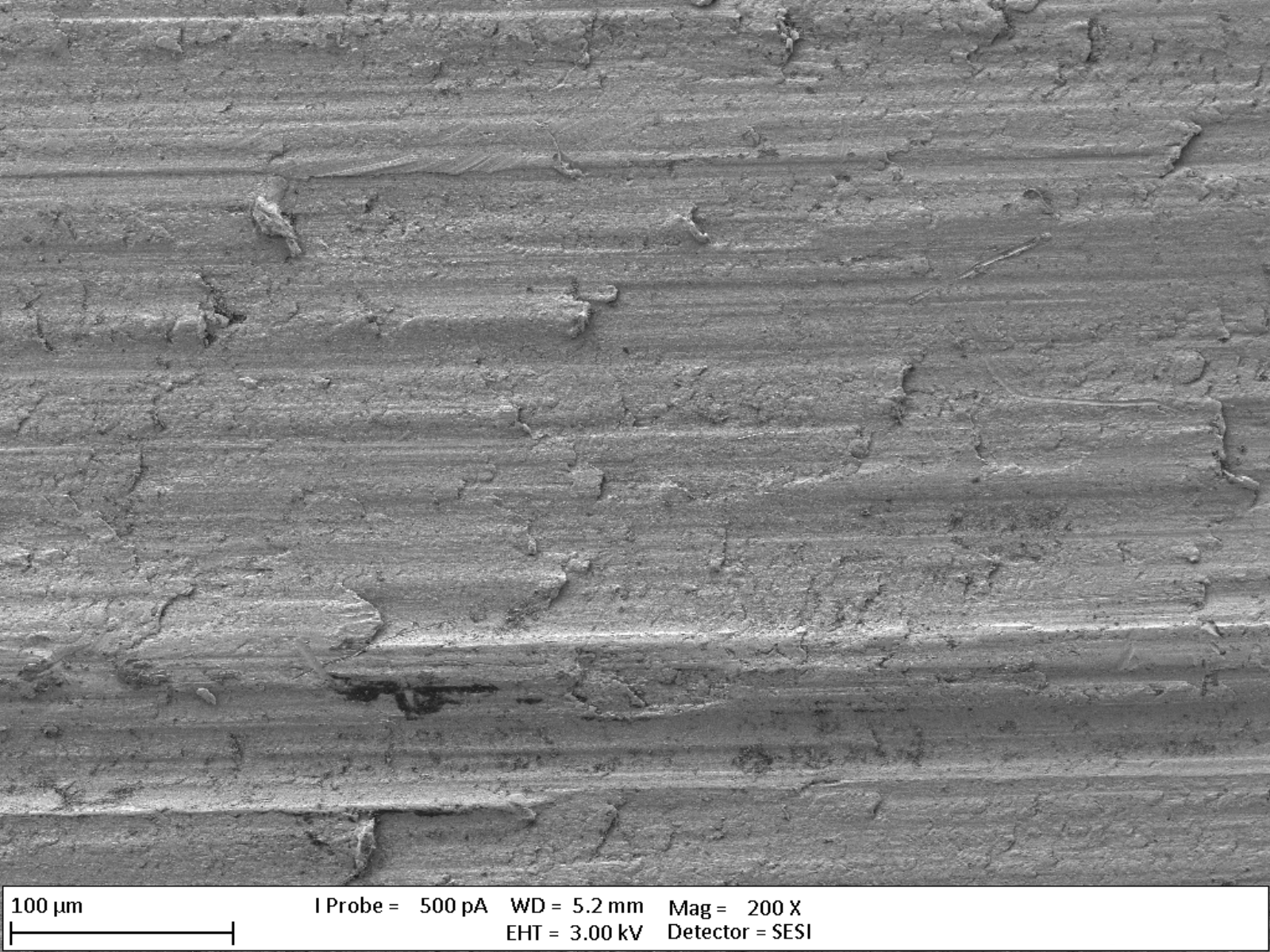}%
\caption{\label{fig:n1}SEM micrograph of the surface of the S355J2+N sample.}%
\end{figure}
\begin{figure}[!h]
\includegraphics[scale=0.4]{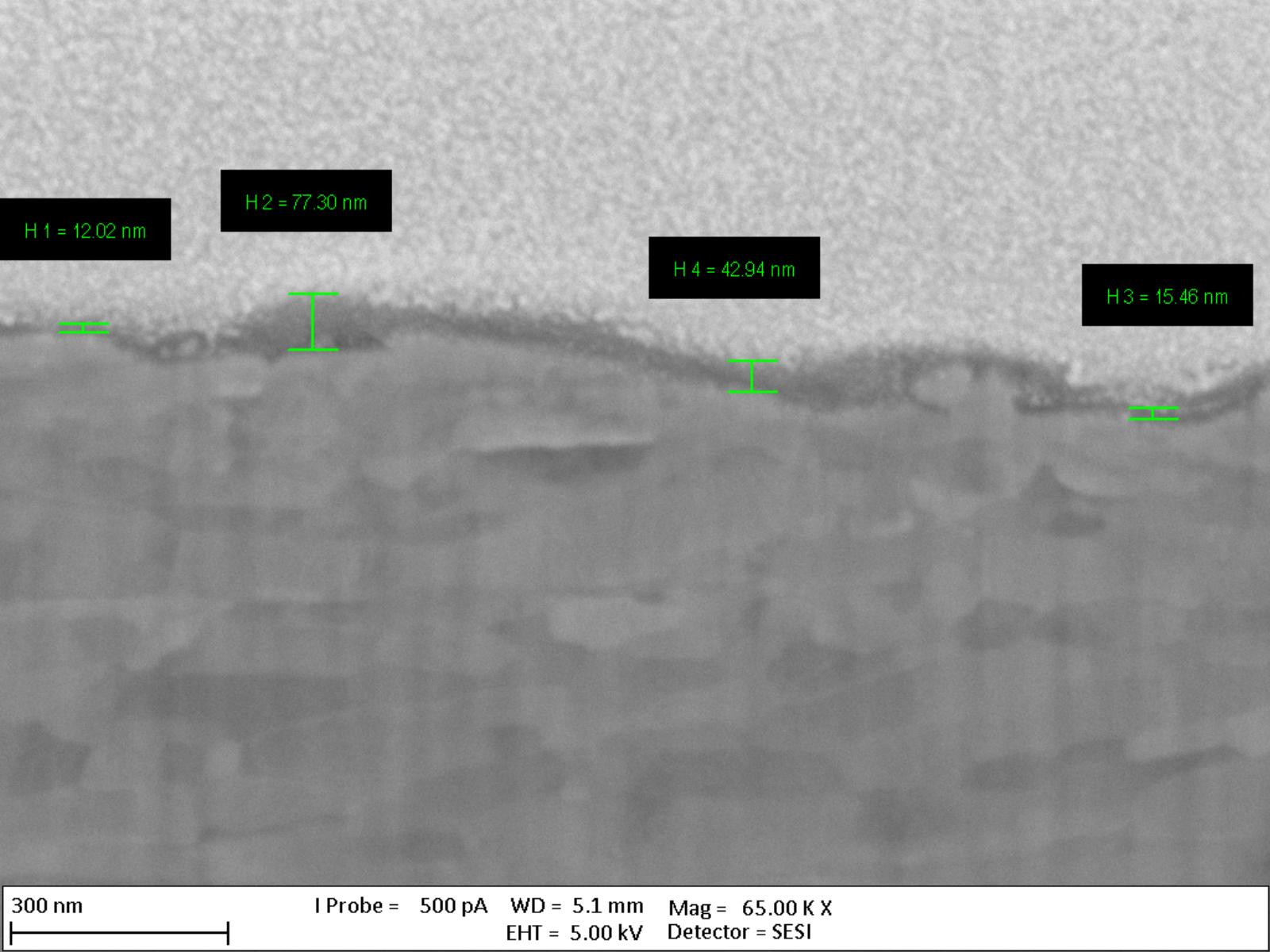}%
\caption{\label{fig:n2}SEM-FIB cross-section micrograph of the S355J2+N surface highlighting an oxidated spot.The thickness of the oxide layer is highlighted by bars in four different positions.}%
\end{figure}
\clearpage
\subsection{\label{sec:xps}X-Ray Photoelectron Spectroscopy}
\begin{figure}
\centering
    \begin{subfigure}{0.8\textwidth}
        \includegraphics[width =\textwidth]{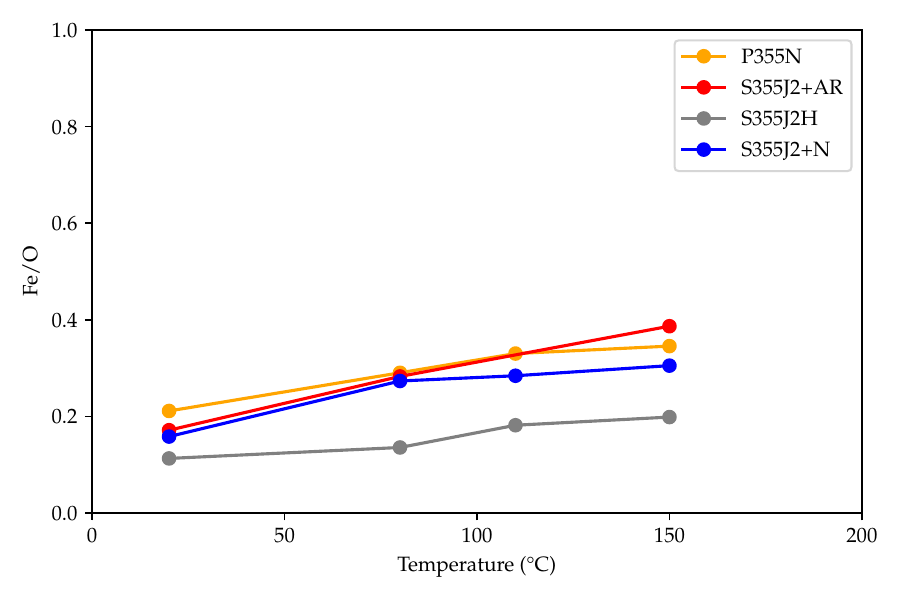}
        \caption{}
        \label{fig:13a}
    \end{subfigure}
    \begin{subfigure}{0.8\textwidth}
        \includegraphics[width = \textwidth]{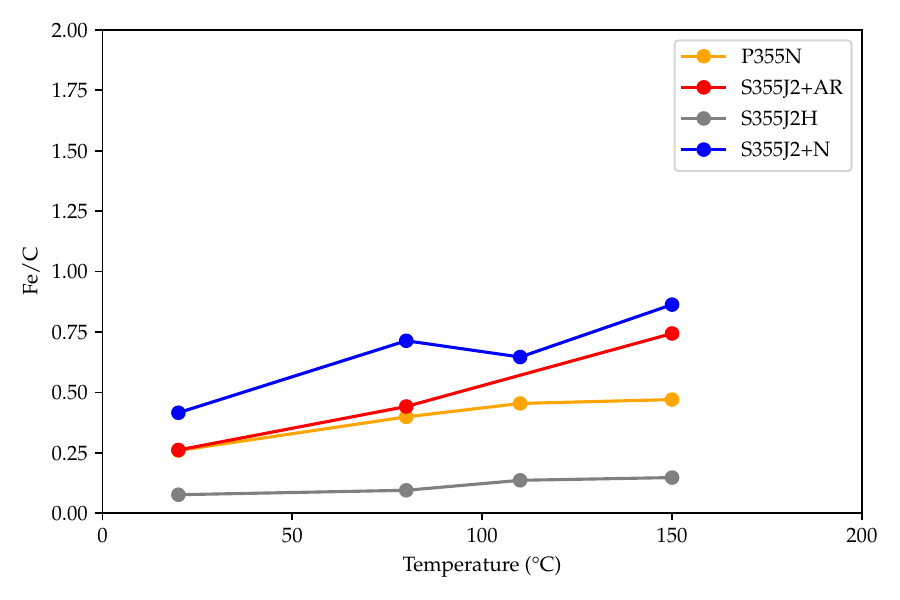}
        \caption{}
        \label{fig:13b}
    \end{subfigure}
\caption{Elemental composition as Fe/O and Fe/C ratios at RT, and after in-situ thermal treatments at 80°C, 110°C, and 150°C.}
\label{fig:13}
\end{figure}

\begin{figure}
\centering
    \begin{subfigure}{0.8\textwidth}
        \includegraphics[width =\textwidth]{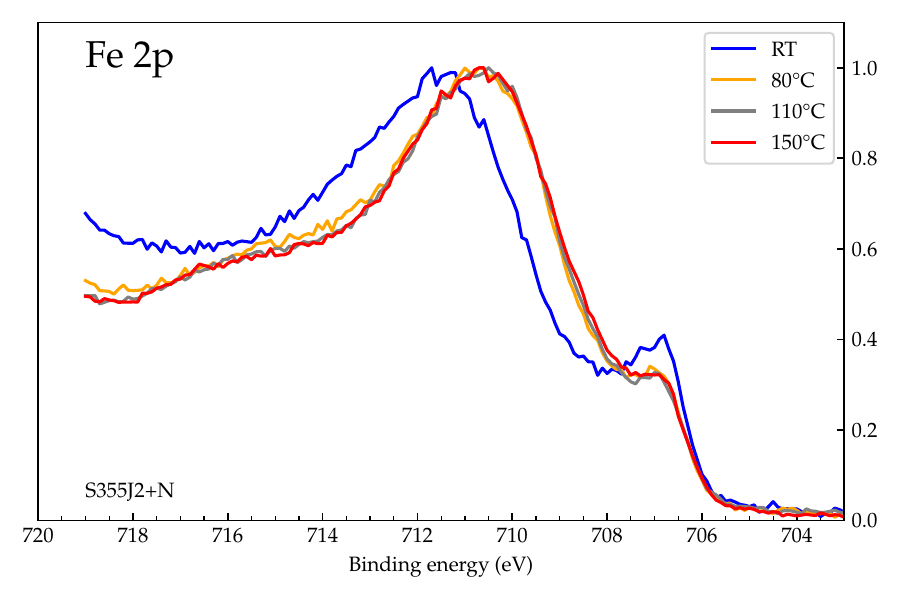}
        \caption{}
        \label{fig:14a}
    \end{subfigure}
    \begin{subfigure}{0.8\textwidth}
        \includegraphics[width = \textwidth]{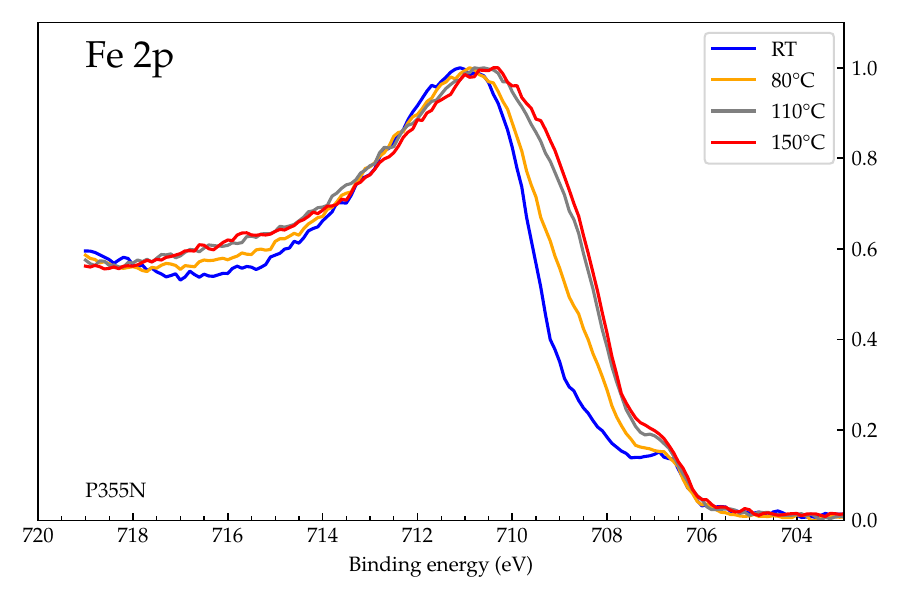}
        \caption{}
        \label{fig:14b}
    \end{subfigure}
\caption{Evolution of the Fe 2p$_{3/2}$ XP spectra between RT and 150°C of two selected samples: (a) S355J2+N and (b) P355N.}
\label{fig:14}
\end{figure}

\begin{figure}
\centering
    \begin{subfigure}{0.8\textwidth}
        \includegraphics[width =\textwidth]{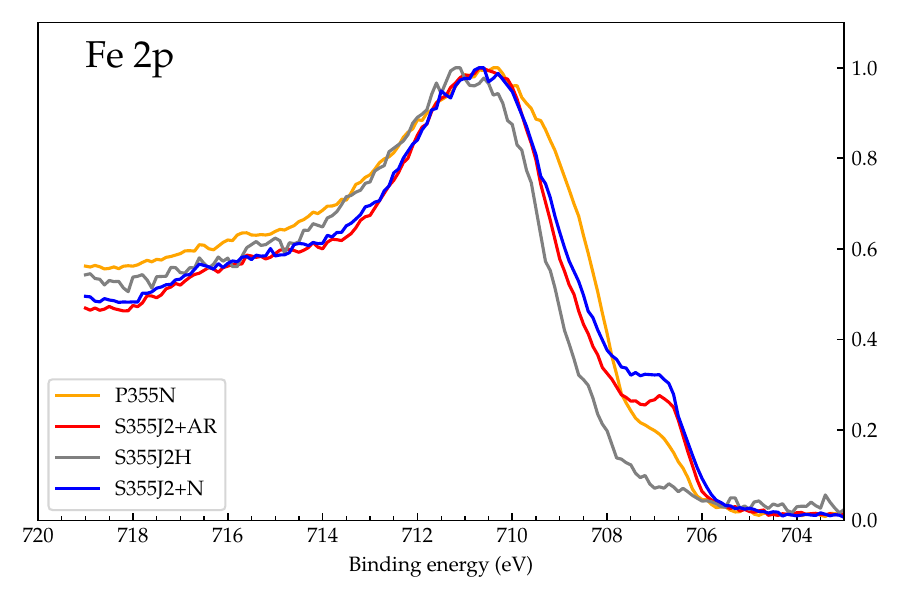}
        \caption{}
        \label{fig:15a}
    \end{subfigure}
    \begin{subfigure}{0.8\textwidth}
        \includegraphics[width = \textwidth]{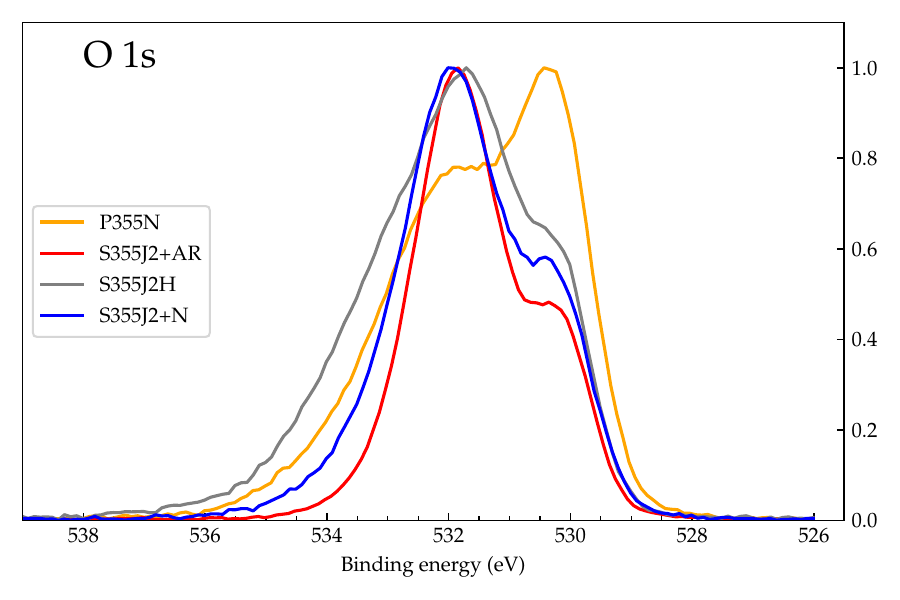}
        \caption{}
        \label{fig:15b}
    \end{subfigure}
\caption{(a) Fe 2p$_{3/2}$ and (b) O 1s XP spectra of the steel samples after thermal treatment at 150°C.}
\label{fig:15}
\end{figure}

The surface chemical composition of the steel samples, investigated by X-ray Photoelectron Spectroscopy (XPS), revealed C, O, and Fe as major constituents, along with a minor content of N, P, Si, and other metals. The evolution of the samples’ surface was also investigated after in situ thermal treatments at 80°C and 150°C.\\
The elemental composition, analysed as the ratios between the three major components (C, O, and Fe) and visualized as Fe/C and Fe/O (\cref{fig:13}), indicated, as a general trend, a progressive increase with the temperature of the relative surface iron content. Over the range of tested temperatures, the S355J2H samples demonstrated a lower relative surface iron content, whereas the other samples had relatively similar results to each other.\\
The detailed Fe 2p spectra probed the evolution of the surface oxides and hydroxides upon heating. At room temperature, the Fe 2p$_{3/2}$ line was found in all samples at binding energies between 711.0 and 711.2 eV, indicating a Fe(III) oxide-hydroxide species (Fe(O)OH)\cite{martino1,martino2}. After prolonged heating at 80°C, a shift to lower binding energies was observed (approximately 710.7–710.8 eV) as a result of a surface rearrangement possibly consistent with a dehydration/dehydroxylation process. Concomitantly, Fe(II) satellites – broad features centred at 715 eV – became apparent and supported the occurrence of Fe(II) species in the surface layer probed by XPS. An increase in the thermal treatment temperature (up to 150°C) did not induce any further transformations on the structural steels (an example in \cref{fig:14a}). On the other hand, above 80°C, the Fe2p lines of P355N further evolved, progressively shifting to lower binding energies (\cref{fig:14b}). At 150°C (\cref{fig:15a}), the P355N steels contained, with respect to the structural steels, a higher fraction of Fe species in oxidation state II, as demonstrated by the region of binding energies centred between 709 and 708.5 eV, characteristic of the Fe3O4 and FeO compounds\cite{martino1,martino2}\\
Moreover, Fe(0) (feature at ca. 706.7 eV) was observed in each sample, indicating a thin oxide layer of few nanometres, except for S355J2H that had, instead, a thicker surface oxide –as also probed by SEM (see \cref{fig:tube2}).\\
The nature of the surface oxide species was also evaluated by O 1s XPS: two components, detected at 532 and 530 eV, were assigned to hydroxides (Fe–OH, terminal or bridging) and bulk oxides (O–Fe–O), respectively\cite{martino1,martino2} (\cref{fig:15b}). Structural steels consisted of a higher relative fraction of hydroxides, differently from the P355N steel that instead had a higher bulk oxides component. It is worth mentioning that a broad O1s component at 533-534 eV, attributed to adsorbed water\cite{martino1,martino2}, was found in the room temperature spectra of the S355J2H, S355J2+N, and P355N steels, with the larger contribution seen in the former. The occurrence of adsorbed water has to be linked to high water outgassing rates; similarly, the high amount of surface hydroxides should indicate a material more prone to physisorb and chemisorb water. These arguments are supported by the higher outgassing value measured for S355JH2, with the presence of adsorbed water, low Fe/O ratio (\cref{fig:13a}) and Fe 2p line shifted toward the hydroxide region (\cref{fig:15}).
\clearpage
\subsection{TEMPERATURE PROGRAMMED DESORPTION}
\begin{figure}[!h]
\includegraphics[scale=1]{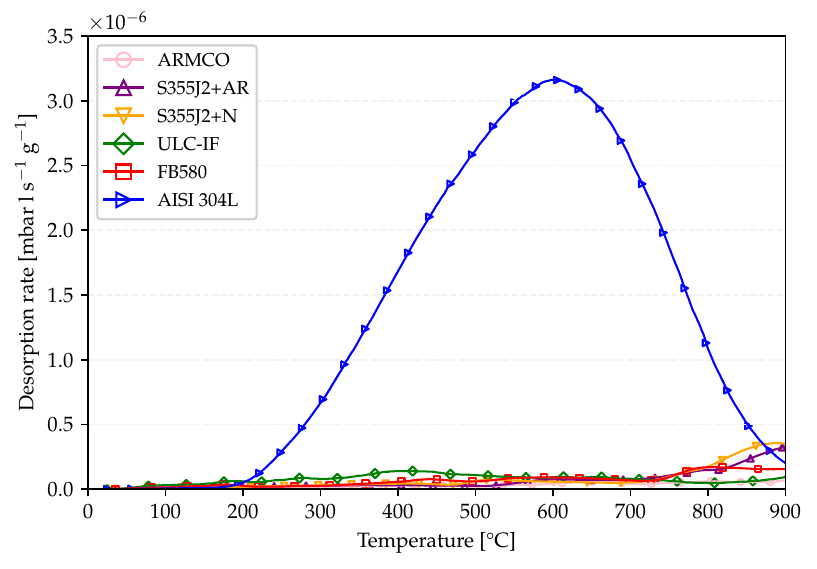}%
\caption{\label{fig:tpdtotal}H$_{2}$ thermal desorption spectra of AISI 304L and flat mild steel samples. The background signal of the TPD system is removed.}%
\end{figure}

\begin{figure}[!h]
\includegraphics[scale=1]{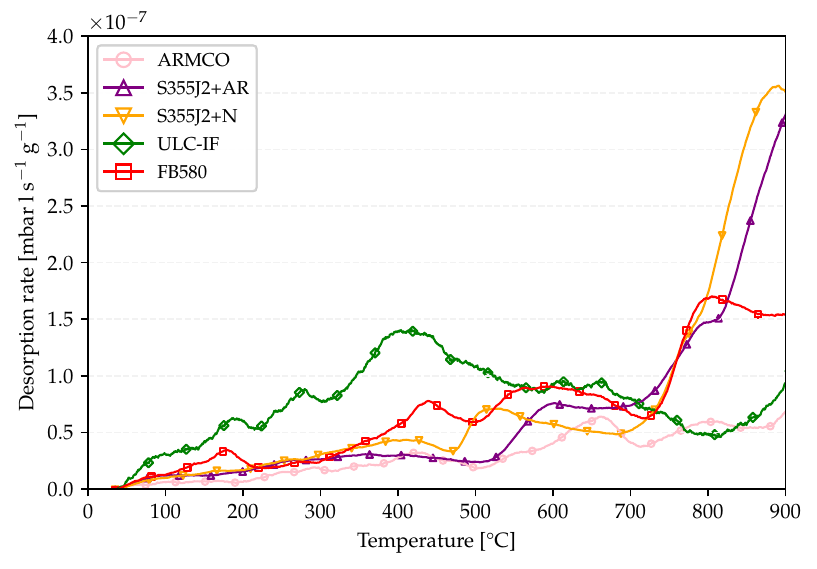}%
\caption{\label{fig:tpdflat}H$_{2}$ thermal desorption spectra of flat mild steel samples (vertical scale reduced x10 with respect to \cref{fig:tpdtotal}). The background signal of the TPD system is removed.}%
\end{figure}

\begin{figure}[!h]
\includegraphics[scale=1]{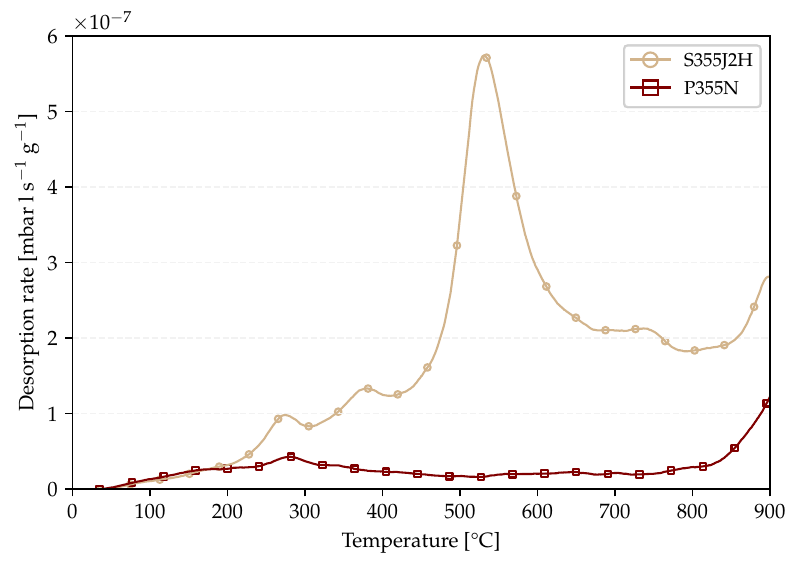}%
\caption{\label{fig:tpdtube}H$_{2}$ thermal desorption spectra of mild steel samples from tubes. The background signal of the TPD system is removed.}%
\end{figure}

\begin{figure}[!h]
\includegraphics[scale=1]{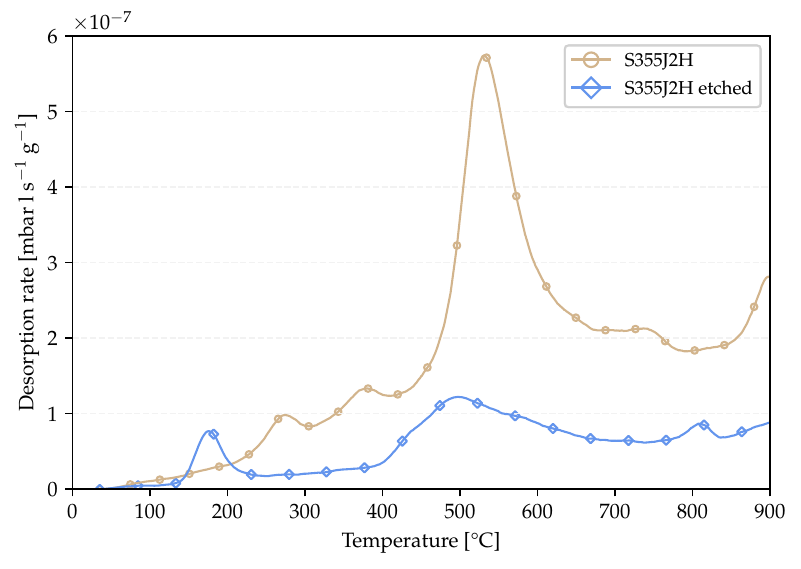}%
\caption{\label{fig:tpdtubetched}H$_{2}$ thermal desorption spectra of S355J2H as received and after etching in a HCl solution. The background signal of the TPD system is removed.}%
\end{figure}

\begin{figure}[!h]
\includegraphics[scale=1]{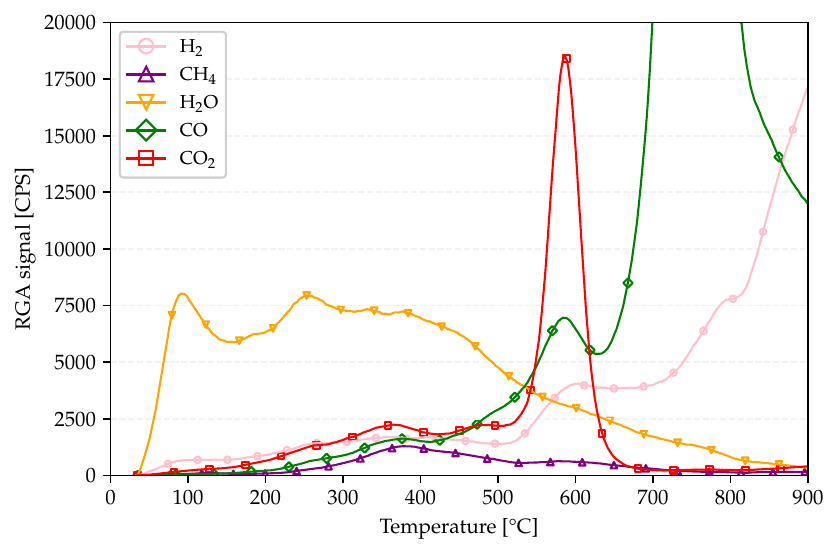}%
\caption{\label{fig:tpdar}Thermal desorption spectra of a S355J2+AR sample. The background signal of the TPD system is removed.}%
\end{figure}

\begin{figure}[!h]
\includegraphics[scale=1]{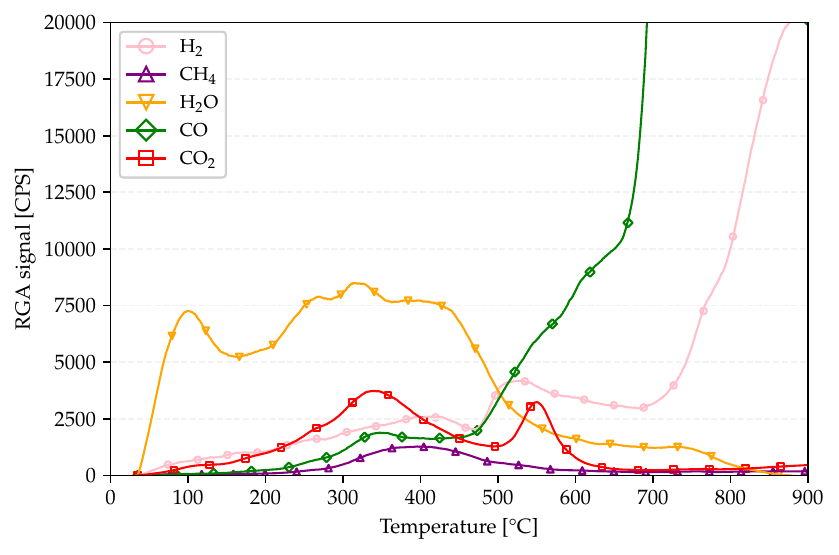}%
\caption{\label{fig:tpdn}Thermal desorption spectra of a S355J2+N sample. The background signal of the TPD system is removed.}%
\end{figure}

The H$_{2}$ thermal desorption spectra, with the background signal removed, are shown in \cref{fig:tpdtotal,fig:tpdflat,fig:tpdtube}. As the thicknesses of the mild steel samples were not the same, the H$_{2}$ desorption spectra are represented normalised to the sample’s weight. This choice can be argued. However, it is justified by the results reported hereafter. All raw data were smoothed for better visualization through a Savitzky-Golay filter\cite{filter} implemented in Python.\\
The signal obtained with AISI 304L samples can be fitted by a Fickian diffusion model that matches the broad peak with a maximum at 615°C (see \cref{fig:tpdtotal}). The obtained diffusion energy is 0.52±0.06 eV, i.e. a typical value for austenitic stainless steels\cite{TDS1,TDS2,TDS3}.\\ 
Among the flat mild steel samples (see \cref{fig:tpdflat}), we can identify a common peak/shoulder around 400°C–440°C. For the S355J2+AR, S355J2+N, ULC-IF, and FB580 samples, an additional concomitant peak/shoulder is observed between 520°C and 580°C. The common presence of a shoulder for the signals of S355J2+AR, S355J2+N, and FB580 is shown again around 780°C.\\ 
Very distinctive and different from the flat samples and from the P355N samples, the latter characterised by a small peak around 270°C and a relatively flat profile, is the H$_{2}$ desorption profile of the S355J2H samples (see \cref{fig:tpdtube}). The desorption signal shows two small shoulders at 270°C and 380°C and a prominent and narrow peak at 530°C. Contrary to all the other types of samples, the signal after the peak starts a descending slope around 740°C before ramping up again at 780°C. To identify the origin of the high peak at 530°C, samples of S355J2H taken from the same delivery batch were immersed in an HCl solution for 2 min to remove the thick oxide layer. Once etched, the samples show a significantly lower desorption signal (\cref{fig:tpdtubetched}). The desorption profile is characterized by a sharp peak at 170°C and a wide peak/shoulder at around 500°C. \\
The H$_{2}$ concentration in the measured samples is calculated by integrating the TPD signal up to 850°C and assuming uniform initial distribution in the volume of the samples. The results of the calculation are reported in \cref{tab:tpd}. The selection of the integration range is arbitrary, justified by the observation that hydrogen released at temperatures above 850°C is tightly bound, rendering it non-diffusible within the temperature range relevant to GWD.  As shown in \cref{fig:tpdtubetched}, a significant quantity of hydrogen can be attributed to the oxide layer. Consequently, the calculated bulk concentrations have to be considered upper limits.\\
The calculated hydrogen concentration in AISI 304L samples is 9.6 to 75 times higher than that evaluated for mild steel samples, the latter showing H contents always below 8 atomic ppm.\\
As expected from their body-centred cubic structure, the mild steel samples have a much lower hydrogen concentration than AISI 304L, which is face-centred cubic austenite. The different thicknesses and geometries of the tested samples do not allow a direct one-to-one comparison between the samples of mild steel; however, some general observations can be drawn. Given the shape of the peaks of the H$_{2}$ desorption profile of the mild steels, it appears that they are primarily influenced by the de-trapping of hydrogen from grain boundaries, lattice defects, carbides, and precipitates rather than Fickian bulk diffusion typical of the austenitic microstructure \cite{trap1,trap2,trap3}. Moreover, the shapes and intensities of the H$_{2}$ peaks are significantly influenced by the presence of thick iron oxides and hydroxides covering the samples. As can be seen from the etching of the S355J2H samples, hydrogen trapped in the oxide layers accounted for almost two-thirds of the total quantity of released hydrogen.\\
As depicted in \cref{fig:tpdar} and \cref{fig:tpdn}, when heating the S355J2+N and S355J2+AR samples within the temperature range of 100°C to 450°C, a correlation becomes apparent between the peaks or shoulders of H$_{2}$ and those of CH$_{4}$, H$_{2}$O, CO, and CO$_{2}$. In this temperature range, water vapour is the leading gas; this result aligns with the XPS measurements (see \cref{sec:xps}) that indicate significant dehydration and dehydroxylation already taking place during a temperature plateau at 80°C. Additionally, an intriguing overlap of peaks or shoulders of H$_{2}$, CO, and CO$_{2}$ is observed between temperatures of 500°C and 590°C. This latter temperature range coincides with the onset of Wustite (FeO) formation through the reduction of Hematite (Fe$_{2}$O$_{3}$) and Magnetite (Fe3O4)\cite{ironoxides}. \\
The shoulders observed between 700°C and 800°C in the hydrogen desorption profile may potentially correspond to phase transformations within the steel samples, specifically from $\alpha$-Fe + Fe$_{3}$C to $\alpha$-Fe + $\gamma$-Fe. Similarly, the steeper shoulders observed between 800°C and 900°C could be attributed to a phase transformation from $\alpha$-Fe + $\gamma$-Fe to $\gamma$-Fe\cite{ironcarbon}. The simultaneous increase in CO desorption further supports these observations.
\begin{table}[h]
\caption{Hydrogen concentration obtained by integrating the TPD spectra up to 850°C. The values reported are the average of at least three samples from the same batch. The background of the TPD system is removed. To convert ppm at. to ppm wt. multiply by 55.85 (molecular weight of iron).}
\begin{tabular}{ccc}
\hline
 Steel& \multicolumn{1}{c}{\begin{tabular}[c]{@{}c@{}}H content\\ {[}ppm at.{]}\end{tabular}} & \multicolumn{1}{c}{\begin{tabular}[c]{@{}c@{}}Thickness\\ {[}cm{]}\end{tabular}} \\
\hline
AISI 304 (as received) & 75 & 0.3\\
S355J2H	& 7.8 & 0.4\\
ULC-IF & 3.7 & 0.69\\
FB580 & 2.8 & 2.1\\
S355J2H etched & 2.7 & 0.4\\
S355J2+AR & 2.0 & 0.3\\
S355J2+N & 1.6 & 0.3\\
ARMCO & 1.2 & 0.2\\
P355N & 1.0 & 0.35\\
\hline
\label{tab:tpd}
\end{tabular}
\end{table}
\clearpage
\subsection{OUTGASSING RATES AFTER BAKEOUT}
The specific outgassing rates for H$_{2}$, CH$_{4}$, CO, and CO$_{2}$ are reported in \cref{tab:accu} and \cref{tab:accu1}. The values were calculated with \cref{eq:accu1} from data obtained by accumulation according to the procedure described in \cref{sec:accu}. The linearity of the pressure during the accumulation has been verified by measuring for different accumulation times.\\
As shown in \cref{tab:accu}, H$_{2}$ is the leading gas, with its outgassing rate being equal to or lower than 7.0$\times$10$^{-14}$ mbar l s$^{-1}$ cm$^{-2}$ after a 48 h bakeout at 80°C. A difference can be observed between the tubes and the flat sample, with the latter showing values below 6$\times$10$^{-15}$ mbar l s$^{-1}$ cm$^{-2}$. The measured specific outgassing rates of CH$_{4}$, CO and CO$_{2}$ range between 3$\times$10$^{-15}$ and 2$\times$10$^{-17}$ mbar l s$^{-1}$ cm$^{-2}$.\\
The following bakeout at 150°C for 48 hours (refer to Table 5 and Figure 16) yielded striking results. While the H$_{2}$ specific outgassing rate of the S355J2H tube remained constant at 1$\times$10$^{-14}$ mbar l s$^{-1}$ cm$^{-2}$, the S355J2+AR and P355N samples exhibited reductions by factors of 2.5 and 4, respectively. Additionally, for all other tested alloys, the total H$_{2}$ outgassing rate fell below the detection limit, defined as 50\% of the background value. As per H$_{2}$, CO and CO$_{2}$ show an important outgassing rate reduction. CH$_{4}$ outgassing rates showed little to no decrease for most of the steel grades when the bakeout temperature was increased from 80°C to 150°C.
The CO specific outgassing rate of the S355J2H tube could not be accurately determined due to a significant virtual leak originating from the sample itself. \cref{fig:virtualeak} illustrates that during the RGA measurement of accumulated gas, the increase in mass-28 signal correlates with rises in mass-14 and mass-40 signals, typical for N$_{2}$ and Ar. The mass-12 signal (C) could not be reliably used for calculating the CO contribution, as it may stem from fragmentation patterns of other gases like CH$_{4}$ and CO$_{2}$. Despite extensive leak detection efforts, no external leaks were found in the vacuum chamber or measuring system. The presence of virtual leaks suggests surface deterioration. \cref{fig:tube2} indicates porosities and cracks in the oxide layer likely responsible for an air in-leakage.
The H$_{2}$ specific outgassing rates are qualitatively consistent with the hydrogen concentration measured by TPD. Compared with austenitic stainless steels, after bakeout at 150°C for 24 hours\cite{chiggiato}, i.e 3$\times$10$^{-12}$ mbar l s$^{-1}$ cm$^{-2}$, the mild steel samples show from two to three orders of magnitude lower H$_{2}$ specific outgassing rates, therefore attaining values measured for a few mm thick AISI 304L after vacuum firing (950°C, 2 h)\cite{chiggiato} or air bakeout (390°C, 100 h)\cite{bernardini}. The hydrogen concentration ratios and the ratios of specific hydrogen outgassing rates in mild steels compared to austenitic steels don't align quantitatively. This mismatch might be because hydrogen in mild steel is mainly trapped within the material's bulk and surface oxide, thus not participating in the diffusion process that causes outgassing at room temperature.
\begin{figure}[!h]
\includegraphics[scale=1]{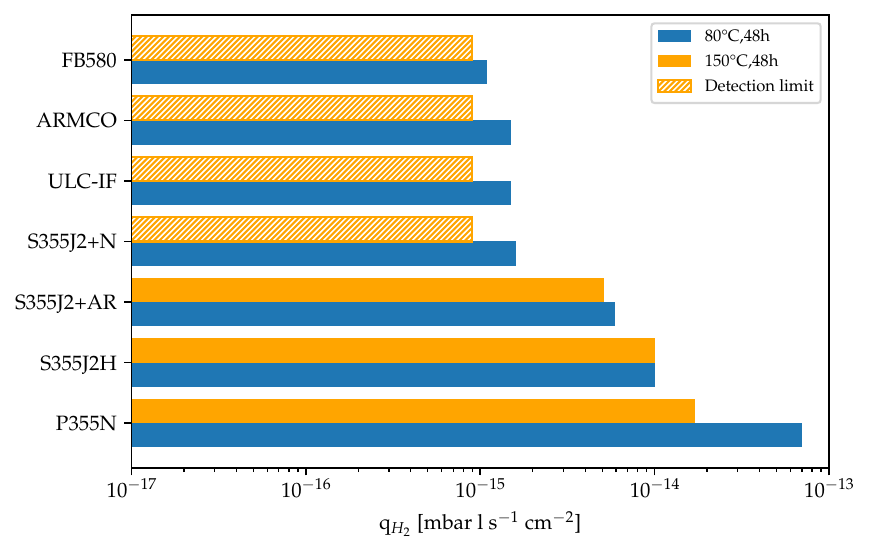}%
\caption{\label{fig:accuH}Comparison of H$_{2}$  specific outgassing rate reported in \cref{tab:accu,tab:accu1}. The system sensitivity (see definition in the text) normalised to the sample surface area is plotted as orange-dashed columns when the measured values are below such a limit.}%
\end{figure}
\begin{figure}[!h]
\includegraphics[scale=1]{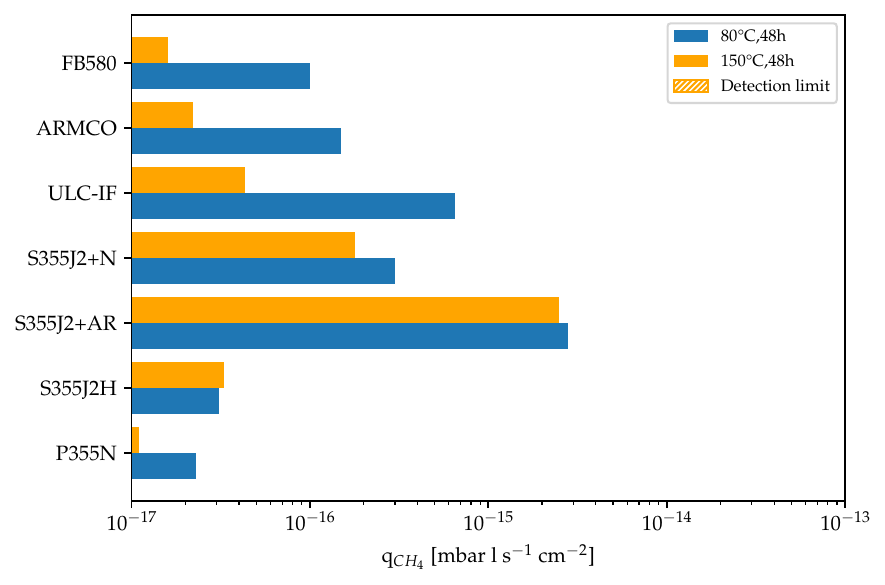}%
\caption{\label{fig:accuCH4}Comparison of CH$_{4}$  specific outgassing rate reported in \cref{tab:accu} and \cref{tab:accu1}. The system sensitivity (see definition in the text) normalised to the sample surface area is plotted as an orange-dashed column when the measured values are below such a limit.}%
\end{figure}
\begin{figure}[!h]
\includegraphics[scale=1]{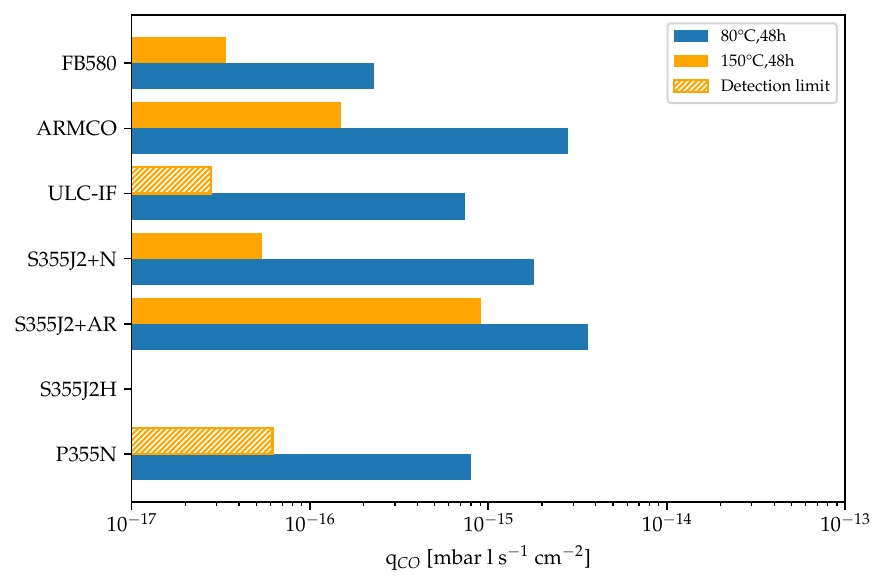}%
\caption{\label{fig:accuCO}Comparison of CO  specific outgassing rate reported in \cref{tab:accu} and \cref{tab:accu1}. The system sensitivity (see definition in the text) normalised to the sample surface area is plotted as an orange-dashed column when the measured values are below such a limit. The lack of the data for S355J2H is explained in the text.}%
\end{figure}
\begin{figure}[!h]
\includegraphics[scale=1]{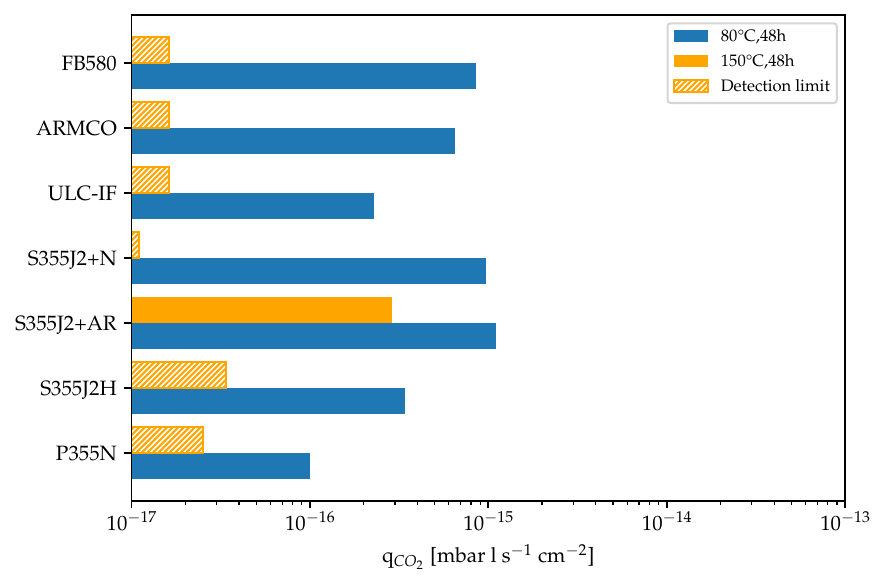}%
\caption{\label{fig:accuCO2}Comparison of CO$_{2}$  specific outgassing rate reported in \cref{tab:accu} and \cref{tab:accu1}.The system sensitivity (see definition in the text) normalised to the sample surface area is plotted as orange-dashed columns when the measured values are below such a limit.}%
\end{figure}
\begin{figure}[!h]
\includegraphics[scale=1]{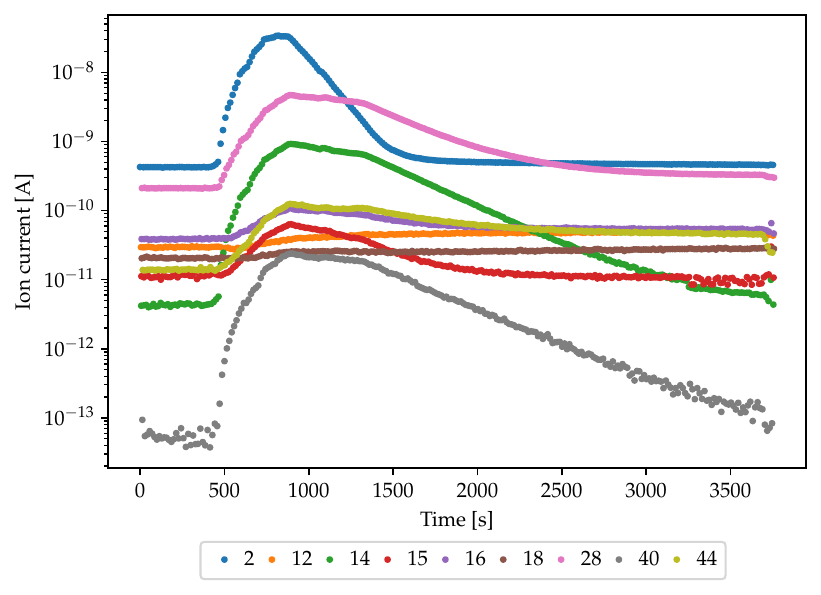}%
\caption{\label{fig:virtualeak} MID scan of the accumulated gas in a  S355J2H tube  after 80°C, 48 h bake-out.}%
\end{figure}
\begin{table}[]
\caption{Specific outgassing rates for the different steels at 21±2°C after bake-out at 80°C for 48 h. Background signal removed.}
\begin{tabular}{|c|cccc|}
\hline
\multirow{2}{*}{Steel grade} & \multicolumn{4}{c|}{\begin{tabular}[c]{@{}c@{}}Specific outgassing rate\\ {[}mbar l s$^{-1}$ cm$^{-2}${]}\end{tabular}} \\ \cline{2-5} 
          & \multicolumn{1}{c|}{H$_{2}$} & \multicolumn{1}{c|}{CH$_{4}$} & \multicolumn{1}{c|}{CO} & CO$_{2}$ \\ \hline
P355N     & \multicolumn{1}{c|}{7.0 $\times$ 10$^{-14}$} & \multicolumn{1}{c|}{2.3 $\times$ 10$^{-17}$}  & \multicolumn{1}{c|}{8.0 $\times$ 10$^{-16}$}  & 1.0 $\times$ 10$^{-16}$   \\ \hline
S355J2H   & \multicolumn{1}{c|}{1.0 $\times$ 10$^{-14}$} & \multicolumn{1}{c|}{3.1 $\times$ 10$^{-17}$}  & \multicolumn{1}{c|}{N/A}  & 3.4 $\times$ 10$^{-16}$   \\ \hline
S355J2+AR & \multicolumn{1}{c|}{5.9 $\times$ 10$^{-15}$}  & \multicolumn{1}{c|}{2.8 $\times$ 10$^{-15}$}   & \multicolumn{1}{c|}{3.6 $\times$ 10$^{-15}$}   &  1.1 $\times$ 10$^{-15}$   \\ \hline
S355J2+N  & \multicolumn{1}{c|}{1.6 $\times$ 10$^{-15}$}  & \multicolumn{1}{c|}{3.0 $\times$ 10$^{-16}$}   & \multicolumn{1}{c|}{1.8 $\times$ 10$^{-15}$}   &   9.7 $\times$ 10$^{-16}$  \\ \hline
ULC-IF    & \multicolumn{1}{c|}{1.5 $\times$ 10$^{-15}$}  & \multicolumn{1}{c|}{6.5 $\times$ 10$^{-16}$}   & \multicolumn{1}{c|}{7.4 $\times$ 10$^{-16}$}   &   2.3 $\times$ 10$^{-16}$  \\ \hline
ARMCO     & \multicolumn{1}{c|}{1.5 $\times$ 10$^{-15}$}  & \multicolumn{1}{c|}{1.5 $\times$ 10$^{-16}$}   & \multicolumn{1}{c|}{2.8 $\times$ 10$^{-15}$}   &   6.5 $\times$ 10$^{-15}$  \\ \hline
FB580     & \multicolumn{1}{c|}{1.1 $\times$ 10$^{-15}$}  & \multicolumn{1}{c|}{1.0 $\times$ 10$^{-16}$}   & \multicolumn{1}{c|}{2.3 $\times$ 10$^{-16}$}   &   8.5 $\times$ 10$^{-16}$  \\ \hline
\end{tabular}
\label{tab:accu}
\end{table}
\begin{table}[]
\caption{Specific outgassing rates for the different steels at 21±2°C after bake-out at 150°C for 48h. Background signal removed. BSS: Below system sensitivity.}
\begin{tabular}{|c|cccc|}
\hline
\multirow{2}{*}{Steel grade} & \multicolumn{4}{c|}{\begin{tabular}[c]{@{}c@{}}Specific outgassing rate\\ {[}mbar l s$^{-1}$ cm$^{-2}${]}\end{tabular}} \\ \cline{2-5} 
          & \multicolumn{1}{c|}{H$_{2}$} & \multicolumn{1}{c|}{CH$_{4}$} & \multicolumn{1}{c|}{CO} & CO$_{2}$ \\ \hline
P355N     & \multicolumn{1}{c|}{1.7 $\times$ 10$^{-14}$} & \multicolumn{1}{c|}{BSS}  & \multicolumn{1}{c|}{BSS}  & BSS   \\ \hline
S355J2H   & \multicolumn{1}{c|}{1.0 $\times$ 10$^{-14}$} & \multicolumn{1}{c|}{3.3 $\times$ 10$^{-17}$}  & \multicolumn{1}{c|}{N/A}  & BSS   \\ \hline
S355J2+AR & \multicolumn{1}{c|}{2.2 $\times$ 10$^{-15}$}  & \multicolumn{1}{c|}{2.5 $\times$ 10$^{-15}$}   & \multicolumn{1}{c|}{9.1 $\times$ 10$^{-16}$}   &  2.9 $\times$ 10$^{-16}$   \\ \hline
S355J2+N  & \multicolumn{1}{c|}{BSS}  & \multicolumn{1}{c|}{1.8 $\times$ 10$^{-16}$}   & \multicolumn{1}{c|}{5.4 $\times$ 10$^{-17}$}   &   BSS \\ \hline
ULC-IF    & \multicolumn{1}{c|}{BSS}  & \multicolumn{1}{c|}{4.3 $\times$ 10$^{-17}$}   & \multicolumn{1}{c|}{BSS}   &   BSS \\ \hline
ARMCO     & \multicolumn{1}{c|}{BSS}  & \multicolumn{1}{c|}{2.2 $\times$ 10$^{-17}$}   & \multicolumn{1}{c|}{1.5 $\times$ 10$^{-16}$}   &   BSS \\ \hline
FB580     & \multicolumn{1}{c|}{BSS}  & \multicolumn{1}{c|}{1.6 $\times$ 10$^{-17}$}   & \multicolumn{1}{c|}{3.4 $\times$ 10$^{-17}$}   &   BSS  \\ \hline
\end{tabular}
\label{tab:accu1}
\end{table}
\clearpage
\section{ULTIMATE PRESSURE AFTER LOW-TEMPERATURE BAKEOUT}
\begin{figure}[!h]
\includegraphics[scale=1]{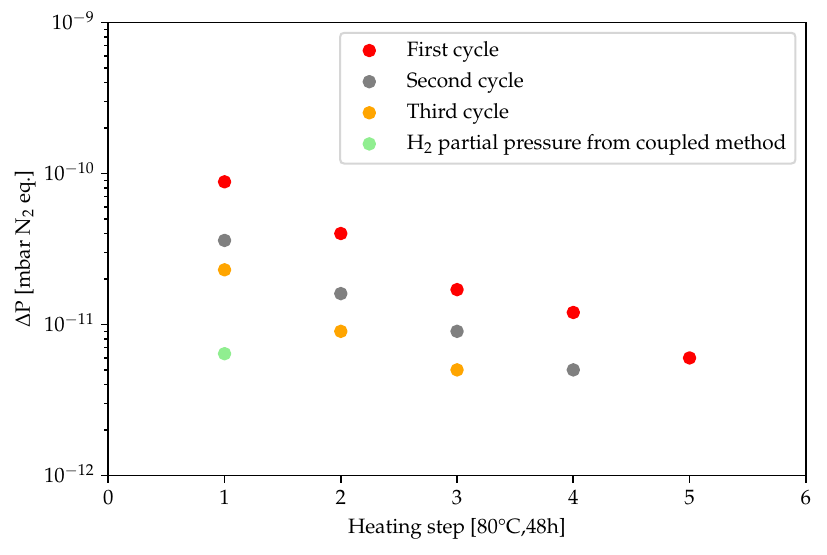}%
\caption{\label{fig:ups1} Ultimate pressure measurements at 21°C. Background removed. The number of cycles has been reduced progressively with heating cycles given the measured values being too close to the background one and thus not allowing the calculation of a significant value. The calculated H$_{2}$ partial pressure resulting from accumulation measurements is also shown at the first heating step.}%
\end{figure}

\begin{figure}[!h]
\includegraphics[scale=1]{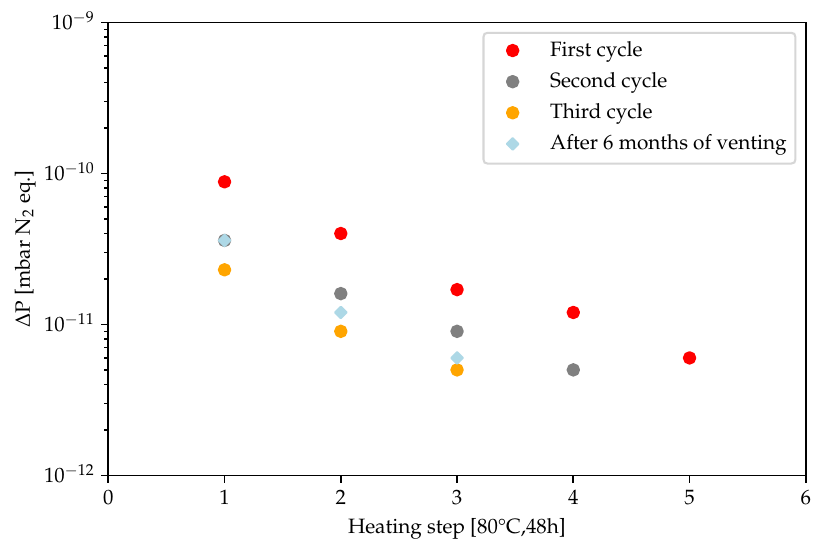}%
\caption{\label{fig:ups2} Ultimate pressure measurements at 21°C including a fourth cycle obtained after a 6-month exposure in air. Background removed.}%
\end{figure}

The ultimate pressure measurements were performed on a P355N vacuum chamber (6.3 cm inner diameter, 320 cm long, 6333 cm$^{2}$ internal surface area). The results, with the background signal removed (equal to 1.5 $\times$ 10$^{-11}$ mbar N$_{2}$ eq.) and expressed in N$_{2}$ equivalent, are reported in \cref{fig:ups1}.\\
The P355N chamber, subjected to a series of consecutive heating steps at 80°C, each lasting 48 h, consistently exhibited a halving of total pressure between each heating step. Based on the results of the initial cycle, it was decided to vent the P355N chamber to laboratory atmosphere conditions (temperature: 21±2°C, relative humidity of about 50\%) for 24 h and proceed with a second cycle of bakeouts to observe if the same pressure reduction trend persisted.\\
As depicted in \cref{fig:ups1}, following the initial heating step of the second cycle, the pressure was 2.4 times lower compared to that of the first cycle. The pressure reduction between steps in the second cycle closely resembled that of the first cycle. After completing four heating steps, the tube was again vented with air at room temperature for an additional 24 h before extending the measurement campaign to a third cycle. The ultimate pressure after the initial heating step between the second and third cycles did not exhibit the same decrease as observed in the first two cycles. However, the decay between the heating steps remained consistent with previous observations.\\
The system was not equipped with an RGA to prevent undue contributions to the background signal, making it impossible to assess the gas species contributing to the total pressure directly. Nonetheless, valuable insights can be obtained from the specific outgassing rates measured using the coupled method for identical materials from the same production batch and subjected to identical cleaning procedures. For such materials, H$_{2}$ emerges as the dominant accumulated gas, while the contributions of CH$_{4}$, CO, and CO$_{2}$ are at least 87 times lower (see \cref{tab:accu}). Assuming the same specific H$_{2}$ outgassing rate for the P355N vacuum chamber under examination, the N$_{2}$ equivalent pressure contribution would be 6.5 $\times$ 10$^{-12}$ mbar. Considering that the difference between the measured pressure and the H$_{2}$ contribution is approximately one order of magnitude, the measured pressure can primarily be attributed to water vapor, considering that its outgassing rate cannot be determined by accumulation methods, as noted previously.\\
After the three cycles, i.e. in total 12 bakeouts at 80°C for 48 h, the specific H$_{2}$ outgassing rate of the P355N vacuum chamber was directly measured by the coupled method following an additional bakeout at 80°C for 48 h. The measured value was 7.5 $\times$ 10$^{-16}$ mbar l s$^{-1}$ cm$^{-2}$, nearly two orders of magnitude lower than the value of \cref{tab:accu}, which was obtained after a single bakeout step at 80°C.\\
Once the outgassing measurement was completed, the P355N vacuum chamber was disconnected from the accumulation system and stored in air on a laboratory’s shelf for 6 months, protected only by plastic caps at the extremities. Afterwards, the same chamber was reinstalled again on the ultimate pressure system, and it underwent a fourth cycle of bakeouts. The objective of the measurement was to test the effect of the storage in air on the ultimate pressure with respect to the values recorded in the previous three cycles.\\ 
After the initial bakeout of the fourth cycle, the ultimate pressure matches that of the second cycle (see \cref{fig:ups2}). Subsequently, a reduction by a factor of three is achieved after an additional bakeout, and a third bakeout results in an ultimate pressure comparable to the latest one measured six months before exposure to air.
The observed behaviours and trends across cycles can be attributed to surface modifications resulting from repeated bakeouts at 80°C. The conditioning effect appears permanent within the specified time frame of air exposure, opening the way for further investigations. 
\clearpage
\section{CONCLUSIONS}
In this work, we explored the UHV compatibility of mild steels and their applicability as structural materials for vacuum tubes of next-generation GWD.\\
For the selected samples, before bakeout, the water vapour outgassing rate follows the usual reciprocal function of the pumping time, while the values are from 1.3 to 9 times higher than those typically measured for ordinary austenitic stainless steel. These differences and large spans could be ascribed to the morphology of the surface oxide layers, whose thickness can range from 1 nm to 10 $\mu$m as highlighted by SEM/FIB cross-sectional analysis. \\
The hydrogen content in mild-steel samples, analysed by TPD up to 850°C, is significantly lower—ranging from 10 to 75 times less—than the levels typically observed in AISI 304L. This disparity was anticipated due to the lower hydrogen solubility inherent in body-centred cubic crystallographic structures, which are characteristic of mild steel. The TPD peaks indicate a more intricate desorption process than the typical behaviour seen in austenitic stainless steels, where hydrogen diffusion emerges as the dominant factor. The shape and positioning of these desorption peaks and the simultaneous release of carbon-containing molecules suggest that hydrogen liberation may be linked to de-trapping from the metal or the oxide layer. Structural transformations enhance the former, while chemical changes in the topmost layer may influence the latter. Notably, XPS detects alterations in the chemical composition of the oxide layer after just an 80°C bakeout. Although the investigation into the exact origin of these peaks falls beyond the scope of this study, given their significant impact on outgassing rates, they will be subject to further examination in subsequent research.\\
The bakeout temperatures chosen for this study are relatively low compared to the typical values used for UHV systems. This decision stems from the most practical and efficient method for baking the beampipe in GWD— utilizing the joule effect with electrical current applied to the vessel’s walls. If the beampipes were constructed from mild steel in a pipeline-like design, the wall thickness would be approximately 1 cm thick, 2.5 to 3 times thicker than those currently utilized in GWD. Assuming the same insulation and bakeout temperature applied for LIGO and VIRGO (15 cm insulation and 150°C) and considering that the electrical resistivity of mild steels is 80\% lower than that of 304L, the current to be applied is around 6000 A. It becomes apparent that bakeout would only be feasible at low temperatures, approximately around 80°C, where the current will drop to 4000 A, assuming the same conditions.
This study demonstrates that such low temperatures do not pose an issue for H$_{2}$ outgassing. Specifically, the measured H$_{2}$ specific outgassing rates, recorded at room temperature following a 48 h bakeout at 80°C, fall within the range of 10$^{-14}$ mbar l s$^{-1}$ cm$^{-2}$, comparable to those observed in vacuum-fired or air-baked austenitic stainless steel vacuum chambers. These values align with the requirements for future gravitational wave detectors and provide the benefit of avoiding expensive and time-consuming high-temperature degassing treatments.
However, the outgassing rate of water vapour remains a concern, as it predominantly influences the ultimate pressure after several days of bakeout at 80°C. Nevertheless, there is a gradual decrease in the ultimate pressure observed after a series of bakeouts at 80°C, indicating a progressive reduction in the outgassing rate of water vapour. This conditioning of the surface appears to be semi-permanent, lasting at least after half a year of exposure to air on a laboratory shelf. The nature of this behaviour and its implications for the feasibility of mild steel beampipes in GWD will be explored in further studies. It is also important to focus on developing surface quality, stability, and corrosion resistance to substantiate the feasibility of mild steel beampipes for the next generation of gravitational wave detectors. 

\begin{acknowledgments}
We thank Mr Rowan Hill-James and Dr Adrienn Baris for their contribution to the outgassing and metallurgical measurements, respectively.\\
This work has been sponsored by the Wolfgang Gentner Programme of the German Federal Ministry of Education and Research (grant no. 13E18CHA)
\end{acknowledgments}

\section*{DATA AVAILABILITY}
The data that support the findings of this study are available from the corresponding author upon reasonable request.
\clearpage
\bibliography{mybib}

\end{document}